\documentclass[letterpaper]{article} % DO NOT CHANGE THIS
\usepackage{aaai2026}  % DO NOT CHANGE THIS
\usepackage{times}  % DO NOT CHANGE THIS
\usepackage{helvet}  % DO NOT CHANGE THIS
\usepackage{courier}  % DO NOT CHANGE THIS
\usepackage[hyphens]{url}  % DO NOT CHANGE THIS
\usepackage{graphicx} % DO NOT CHANGE THIS
\usepackage{natbib}  % DO NOT CHANGE THIS AND DO NOT ADD ANY OPTIONS TO IT
\usepackage{caption} % DO NOT CHANGE THIS AND DO NOT ADD ANY OPTIONS TO IT
\usepackage{algorithm}
\usepackage{algorithmic}

\usepackage{newfloat}
\usepackage{listings}
\DeclareCaptionStyle{ruled}{labelfont=normalfont,labelsep=colon,strut=off} % DO NOT CHANGE THIS
\floatstyle{ruled}
\newfloat{listing}{tb}{lst}{}
\floatname{listing}{Listing}
\usepackage{subfig}
\usepackage{multirow}
\usepackage{pifont}
\usepackage{bm}
\newcommand{\figscale}{0.4}

\usepackage{amsmath}
\usepackage{amssymb}
\usepackage{mathtools}
\usepackage{amsthm}
\usepackage{cleveref}
\usepackage{soul}
\usepackage{booktabs}
\theoremstyle{plain}
\newtheorem{theorem}{Theorem}[section]

\newtheorem{lemma}[theorem]{Lemma}

\theoremstyle{definition}

\theoremstyle{remark}

\title{Faster Game Solving via Hyperparameter Schedules}
\author{
    Naifeng Zhang\textsuperscript{\rm 1},
    Stephen Marcus McAleer\textsuperscript{\rm 2},
    Tuomas Sandholm\textsuperscript{\rm 1,3,4,5}
}
\affiliations{
    \textsuperscript{\rm 1}Carnegie Mellon University \\
    \textsuperscript{\rm 2}Anthropic \\
    \textsuperscript{\rm 3}Strategy Robot, Inc. \\
    \textsuperscript{\rm 4}Strategic Machine, Inc. \\
    \textsuperscript{\rm 5}Optimized Markets, Inc. \\
    naifengz@cmu.edu, 
    mcaleer.stephen@gmail.com,
    sandholm@cs.cmu.edu
}

\begin{document}

\maketitle

\begin{abstract}
\textit{Counterfactual regret minimization (CFR)} algorithms are a foundational class of methods for solving imperfect-information games, with the time average of their iterates converging to a Nash equilibrium in two-player zero-sum games. Prior state-of-the-art variants, \textit{Discounted CFR (DCFR)} and \textit{Predictive CFR$^+$ (PCFR$^+$)}, achieved the fastest known practical performance by improving convergence rates over vanilla CFR through discounting early iterations with a fixed discounting scheme. More recently, \textit{Dynamic DCFR (DDCFR)} introduced agent-learned dynamic discounting schemes to further accelerate convergence, at the cost of increased complexity.
To address this, we propose \textit{Hyperparameter Schedules (HSs)}, a remarkably simple, training-free framework that dynamically adjusts CFR discounting over time. HSs aggressively downweight early updates and gradually transition to trusting late-stage strategies, leading to substantially faster convergence with only a few lines of code modifications. We show that HSs derived from just three small extensive-form games generalize effectively to 17 diverse games (including large-scale realistic poker) in both extensive-form and normal-form settings, without any game-specific tuning. Our method establishes a new state of the art for solving two-player zero-sum games.
\end{abstract}

\begin{links}
    \link{Extended version}{https://arxiv.org/pdf/2404.09097}
\end{links}

\section{Introduction}
\label{sec:intro}

Most real-world settings are \textit{imperfect-information games (IIGs)}, where players hold private information and there may also be information that no player knows. To perform well in IIGs, players typically randomize their strategies to avoid being too predictable and to prevent revealing too much of their private information. In many IIGs, this amounts to understanding deception and being able to deceive, challenges that do not arise in perfect-information games such as chess or Go. Over the past two decades, there have been tremendous breakthroughs in techniques for solving imperfect-information games. 
Like most of the literature, we focus on solving two-player zero-sum IIGs by converging to a Nash equilibrium~\cite{nash1950equilibrium}, in which no player can improve by deviating from the equilibrium.
The most popular computational approach for converging to a Nash equilibrium in IIGs is the family of \textit{counterfactual regret minimization (CFR)} algorithms~\cite{zinkevich2007regret}. These methods systematically reduce the regrets of both players through an iterative process, progressively guiding the time-averaged strategy of each player toward a Nash equilibrium (minmax) strategy. CFR-type algorithms serve as the cornerstone for various \textit{state-of-the-art (SoTA)} algorithms in the field~\cite{bowling2015heads,moravvcik2017deepstack,brown2018superhuman,brown2019solving,brown2019superhuman,zarick2020unlocking,farina2021faster,mcaleer2023escher,ddcfr}.

Several innovative improvements to CFR exhibit substantially faster convergence. The introduction of CFR$^+$~\cite{tammelin2014solving} marked a major milestone, demonstrating an order of magnitude better convergence rate than vanilla CFR in practice. Specifically, CFR$^+$ 1) alternates strategy updates between players, 2) uses \textit{regret-matching$^+$ (RM$^+$)} instead of \textit{regret matching (RM)} as its regret minimizer, and 3) implements a linear discounting scheme where the contribution of iteration $t$ to the average strategy is weighted by $t$. 
This advancement played a pivotal role in almost optimally solving the two-player limit Texas hold 'em poker~\cite{bowling2015heads}. Building on the idea of discounting, \textit{Discounted CFR (DCFR)} introduces a family of algorithms that, prior to the present paper, represented the practical SoTA for poker games and other games with poker-like structure~\cite{brown2019solving}. 
Specifically, DCFR incorporates discounts for both regrets and the average strategy by assigning more weight to later iterations using three hyperparameters: $\alpha$, $\beta$, and $\gamma$. 
More recently, \textit{Predictive CFR$^+$ (PCFR$^+$)}~\cite{farina2021faster} stands out as an effective approach that utilizes predictive Blackwell approachability. Prior to the present paper, PCFR$^+$ was the SoTA for zero-sum games other than poker.

Despite the faster convergence achieved by fixed discounting schemes in these CFR variants, recent work shows that dynamically adjusting the discounting scheme across iterations can further accelerate convergence.
Greedy Weights~\cite{zhang2022equilibrium} was the first proposed regret minimization algorithm that dynamically adjusts iteration weights based on runtime observations. The method was primarily focused on approximating equilibria in normal-form games and exhibited poor performance when applied to extensive-form games~\cite{ddcfr}. \textit{Dynamic DCFR (DDCFR)}~\cite{ddcfr}, a recent innovation, uses a \textit{reinforcement learning (RL)} framework trained on four small games to acquire a well-performing discounting scheme that dynamically adjusts the hyperparameters of DCFR at runtime. DDCFR has demonstrated its effectiveness across several games, leveraging the knowledge gained from the training process and game-specific inference. However, the approach incurs additional computation costs from feature calculations, policy training, and network inference for real-time computation of discounting weights. It also assumes that the hyperparameter-changing policy learned in the training games generalizes well to the target games.

In this work, we investigate whether a lightweight and training-free approach can match or even surpass prior SoTA performance by leveraging the benefits of a dynamic discounting scheme without incurring the complexity and computational overhead of training and running an RL agent.
We build on an established insight that, within the CFR family of algorithms, the computed strategy becomes increasingly refined over time, making it reasonable to assign greater weight to more recent iterations. While this idea has been explored in earlier work, we show that the new schemes introduced in this work are significantly more effective than existing ones in the literature. 
We introduce the concept of \textit{Hyperparameter Schedules (HSs)}---schemes that govern how the hyperparameters of the underlying equilibrium-finding algorithm change across iterations. We then propose two performant HSs that, unlike prior work, \textit{aggressively} discount early iterations and \textit{gradually} increase the contribution of later updates as CFR progresses.
Specifically, we augment DCFR with schedules for its hyperparameters ($\alpha$, $\beta$, and $\gamma$) using fewer than 15 lines of code changes, resulting in the algorithm HS-DCFR. Similarly, we enhance PCFR$^+$ by applying a schedule to its hyperparameter $\gamma$ with fewer than 10 lines of code changes, resulting in the algorithm HS-PCFR$^+$. 
We prove that both algorithms converge to a Nash equilibrium at the stated worst-case rates, provided that the hyperparameters of the HS stay within certain ranges. We show via extensive experiments that the new algorithms yield better solutions for a given amount of run time---\textit{in many games by many orders of magnitude without any game-specific tuning}. These algorithms constitute the new SoTA for both extensive-form and normal-form zero-sum games.

\section{Background}

In this section, we introduce the preliminaries of extensive-form (i.e., tree-form) games, which constitute the primary subject of our study. Our notation and presentation follow established conventions in prior work~\cite{brown2015regret,brown2019solving,ddcfr}. 

\subsection{Notation}

Let us denote the finite set of players in an extensive-form game~\cite{osborne1994course} as $\mathcal{N}$, along with a unique player called \textit{chance}, denoted by $c$. Chance has a fixed, known stochastic strategy. A \textit{history} $h$ contains all information at the current situation. The set of all histories in the game tree is denoted by $\mathcal{H}$. 
$\mathcal{A}(h)$ denotes the \textit{actions} available at a given history $h$, and $\mathcal{P}(h)$ is the unique \textit{player} whose turn it is to act at history $h$. 
We write $h \cdot a = h'$ if action $a \in \mathcal{A}(h)$ leads from history $h$ to another history $h'$. 
At \textit{terminal histories} $\mathcal{Z} \subseteq \mathcal{H}$, no actions are available. 
For terminal history $z \in \mathcal{Z}$, each player $i \in \mathcal{N}$ has a utility function $u_i(z) : \mathcal{Z} \to \mathbb{R}$. 
Let us denote the \textit{range of utilities} reachable by player $i$ as $\Delta_i$. Formally, $\Delta_i = \max_{z \in \mathcal{Z}}u_i(z) - \min_{z \in \mathcal{Z}}u_i(z)$ and $\Delta = \max_{i \in \mathcal{N}} \Delta_i$. 

Players not having complete knowledge regarding the game state is represented by \textit{information sets} $\mathcal{I}_i$ for each player $i \in \mathcal{N}$. Any history $h, h'$ in the same information set $I \in \mathcal{I}_i$ are indistinguishable to player $i$. Every non-terminal history $h \in \mathcal{H}$ belongs to exactly one $I$ for each player. 
We define $\mathcal{A}(I)$ as the set of actions such that $\forall h \in I, \mathcal{A}(I) = \mathcal{A}(h)$. Let $|\mathcal{A}_i| = \max_{I \in \mathcal{I}_i} |\mathcal{A}(I)|$, $|\mathcal{A}| = \max_i |\mathcal{A}_i|$, and $|\mathcal{I}| = \sum_{i \in \mathcal{N}} |\mathcal{I}_i|$. 

For player $i$, a \textit{strategy} $\sigma_i(I)$ is a probability distribution on the actions available in each information set $I$. The probability distribution of an action $a$ selected by player $i$ is denoted by $\sigma_i(I, a)$. Given that all histories within an information set $I$ are indistinguishable for player $i$, strategies for all histories within $I$ are identical. Thus, we define $\sigma_i(h, a) = \sigma_i (I, a)$ for any $h \in I$ and $i = \mathcal{P}(h)$.
Let $\sigma_i$ denote a comprehensive strategy for player $i$, which specifies the actions player $i$ chooses in each information set $I\in\mathcal{I}_i$. We define $\sigma_{-i}$ as the strategies of all players other than player $i$. $u_i(\sigma_i, \sigma_{-i})$ is the \textit{expected utility} for player $i$ if player $i$ follows $\sigma_i$ and all other players follow $\sigma_{-i}$.

Let $\sigma$ denote a \textit{strategy profile} that contains a strategy $\sigma_i$ for each player $i$. Let $g \sqsubseteq h$ denote that $g$ is either equal to $h$ or $g$ is a \textit{prefix} of $h$ in the tree structure. 
The \textit{history reach probability} of a history $h$, defined as $\pi^\sigma(h)=\prod_{h^{\prime} \cdot a \sqsubseteq h} \sigma_{\mathcal{P}\left(h^{\prime}\right)}(h^{\prime}, a)$, represents the joint probability of reaching $h$ if all players play according to $\sigma$. The contribution of player $i$ to the history reach probability is defined as $\pi^\sigma_i(h)$ and the contribution of all other players is defined as $\pi^\sigma_{-i}(h)$. 
The \textit{information set reach probability} $\pi^\sigma(I)$ is defined as $\sum_{h \in I}\pi^\sigma(h)$.  
The \textit{interval history reach probability} from history $h$ to $h'$ is expressed as $\pi^\sigma(h, h') = \pi^\sigma(h') / \pi^\sigma(h)$ if $h \sqsubseteq h'$.

\subsection{Nash Equilibrium}

The \textit{best response} to a strategy $\sigma_{-i}$ is any strategy $\operatorname{BR}(\sigma_{-i})$ that maximizes the expected utility for player $i$, such that $u_i(\operatorname{BR}(\sigma_{-i}),\sigma_{-i})=\max_{\sigma_{i}^{\prime}} u_i(\sigma^\prime_{i}, \sigma_{-i})$.
A \textit{Nash equilibrium}~\cite{nash1950equilibrium} is a strategy profile $\sigma^* = (\sigma^*_i, \sigma^*_{-i})$ where each player adopts a strategy that serves as the best response to the strategies chosen by the other players. Formally, $\forall i \in \mathcal{N}, u_i(\sigma_i^*, \sigma_{-i}^*)=\max _{\sigma_i^{\prime}} u_i(\sigma_i^{\prime}, \sigma_{-i}^*)$. 
The \textit{exploitability} of a strategy $\sigma_i$ measures how far it is from a Nash equilibrium in the sense of utility: $e(\sigma_i) = u_i(\sigma^*_i, \operatorname{BR}(\sigma^*_i))-u_i(\sigma_i,\operatorname{BR}(\sigma_i))$. In an \textit{$\epsilon$-Nash equilibrium}, no player has exploitability higher than $\epsilon$. Let $e(\sigma)$ denote the exploitability of a strategy profile. We can compute $e(\sigma)$ using the formula: $e(\sigma) = \sum_{i \in \mathcal{N}} e(\sigma_i) / |\mathcal{N}|$. This is a measure of how far, strategically speaking, a strategy profile is from a Nash equilibrium. 

\subsection{Counterfactual Regret Minimization}

CFR systematically refines players' average strategies by minimizing their regrets over iterations, serving as a prominent equilibrium-finding algorithm for extensive-form IIGs~\cite{zinkevich2007regret}. 
In CFR, the \textit{counterfactual value} represents the expected utility of a given information set $I$ for player $i$ assuming that player $i$ attempts to reach $I$.
Formally, given strategy profile $\sigma$, the counterfactual value for player $i$ at an information set $I \in \mathcal{I}_i$ is defined as $v_i^\sigma(I) = \sum_{h \in I}(\pi_{-i}^\sigma(h) \sum_{z \in \mathcal{Z}}(\pi^\sigma(h, z) u_i(z)))$. 
Similarly, the counterfactual value for player $i$ with action $a$ at $I$ is defined as 
$v_i^\sigma(I, a)=\sum_{h \in I}(\pi_{-i}^\sigma(h) \sum_{z \in \mathcal{Z}}(\pi^\sigma(h \cdot a, z) u_i(z)))$.
Let $\sigma^t$ be the strategy profile on iteration $t$.  
The \textit{instantaneous regret} on iteration $t$ for player $i$ for not choosing action $a$ in $I$ is defined as $r_i^t(I, a)=v_i^{\sigma^t}(I, a)-v_i^{\sigma^t}(I)$. The \textit{cumulative regret} on iteration $T$ for player $i$ of not choosing action $a$ in $I$ is then defined as $R^T_i(I, a) = \sum_{t=1}^T r^t_i (I, a)$. 
We also define $R_i^{T,+}(I, a)=\max (R_i^T(I, a), 0)$ and $R_i^{T}(I)=\max_a (R_i^{T,+}(I, a))$.

CFR typically uses \textit{regret matching (RM)} as the regret minimizer due to its simplicity and lack of parameters~\cite{hart2000simple,cesa2006prediction}. In RM, a player chooses actions in each information set according to a distribution proportional to the positive regret associated with those actions. Formally, on iteration $T+1$, player $i$ chooses actions $a \in \mathcal{A}(I)$ according to the strategy
\begin{equation}
\begin{aligned}
\resizebox{0.9\columnwidth}{!}{$
    \sigma_i^{T+1}(I, a) =
    \left\{
    \begin{array}{ll}
        \frac{R_i^{T,+}(I, a)}{\sum_{a^{\prime} \in \mathcal{A}(I)} R_i^{T,+}\left(I, a^{\prime}\right)}, & \text { if } \sum_{a^{\prime}} R_i^{T,+}\left(I, a^{\prime}\right)>0, \\
        \frac{1}{|\mathcal{A}(I)|}, & \text { otherwise. }
    \end{array}\right.
$}
\end{aligned}
\end{equation}
The \textit{cumulative strategy} on iteration $T$ for player $i$ with action $a$ in an information set $I$ is defined as $C_i^T(I, a)=\sum_{t=1}^T(\pi_i^{\sigma^t}(I) \sigma_i^t(I, a))$.
If player $i$ employs CFR on each iteration, then on iteration $T$, the regret for an information set $I$ satisfies $R_i^T(I) \leq \Delta_i \sqrt{|\mathcal{A}(I)|}\sqrt{T}$. The regret for player $i$ in the entire game satisfies that $R_i^T \leq \sum_{I \in \mathcal{I}_i} R_i^T(I) \leq\left|\mathcal{I}_i\right| \Delta_i \sqrt{\left|\mathcal{A}_i\right|} \sqrt{T}$.
Thus, as $T \rightarrow \infty$, $R_i^T/T \rightarrow 0$. In two-player zero-sum IIGs, if the average regret $R_i^T/T$ of both players is less than or equal to $\epsilon$, their average strategies constitute a $2\epsilon$-equilibrium~\cite{waugh2009abstraction}. Consequently, CFR serves as an anytime algorithm capable of finding an $\epsilon$-Nash equilibrium in zero-sum games.

\subsection{Discounted CFR Variants}

Numerous improvements to CFR have been introduced that significantly enhance convergence speed. Due to space constraints, in this section, we provide only a brief overview of the SoTA discounted CFR algorithms and refer readers to the cited works for more details.

Building on vanilla CFR, CFR$^+$~\cite{tammelin2014solving} improves convergence rate by an order of magnitude in practice via incorporating alternating updates, using a variation of RM, and applying a linear discounting scheme when computing the average strategy.
\textit{Discounted CFR (DCFR)}~\cite{brown2019solving} comprises a family of algorithms that apply discounts to both cumulative regrets and the average strategy. Although DCFR has a theoretical convergence bound that is worse than that of CFR by a constant factor, it has been shown to converge much faster in practice. DCFR's discounting scheme is parameterized by three hyperparameters: $\alpha$, $\beta$, and $\gamma$. 
On iteration $t+1$, DCFR multiplies positive cumulative regrets by $\frac{t^\alpha}{t^\alpha+1}$, negative cumulative regrets by $\frac{t^\beta}{t^\beta+1}$, and contributions to the average strategy by $(\frac{t}{t+1})^\gamma$. Formally, for player $i$,
\begin{equation}
\label{eq:dcfr_alpha_beta}
\begin{aligned}
\resizebox{0.9\columnwidth}{!}{$
    R_i^{t+1}(I, a) =
    \begin{cases}R_i^{t}(I, a) \frac{t^\alpha}{t^\alpha+1}+r_i^{t+1}(I, a), \;\;\text{if } R_i^{t}(I, a)>0, \\
    R_i^{t}(I, a) \frac{t^\beta}{t^\beta+1}+r_i^{t+1}(I, a), \;\;\text {otherwise, and}\end{cases}
$}
\end{aligned}
\end{equation}
\begin{equation}
\label{eq:dcfr_gamma}
\resizebox{0.8\columnwidth}{!}{$
    \hspace{-.07in} C_i^{t+1}(I, a) = C_i^{t}(I, a) (\frac{t}{t+1})^\gamma+\pi_i^{\sigma^{t+1}}(I) \sigma_i^{t+1}(I, a).
$}
\end{equation}
The authors of DCFR recommend setting $(\alpha,\beta,\gamma)$ to $(1.5,0,2)$, as this configuration consistently outperforms CFR$^+$ in their experiments. 

Recently, researchers introduced \textit{Dynamic DCFR (DDCFR)}~\cite{ddcfr} that employs an RL framework. This framework encapsulates CFR's iteration process within an environment, treating the discounting scheme as an agent interacting with it. The RL agent within DDCFR is trained on four small games and, for each test game, DDCFR performs inference on how to dynamically change the hyperparameters $(\alpha,\beta,\gamma)$ of DCFR across iterations. 
\textit{Predictive CFR$^+$ (PCFR$^+$)}~\cite{farina2021faster} extends Blackwell approachability~\cite{blackwell1956analog} to regret minimization to form \textit{predictive regret matching (PRM+)}. PRM+ uses the observed utility in each iteration during the run as the prediction of the next utility. 
Additionally, PCFR$^+$ employs a quadratic discounting scheme that can be formalized in the same way as in DCFR (i.e., Equation~\ref{eq:dcfr_gamma}) by setting $\gamma$ to $2$. 
The authors of DDCFR~\cite{ddcfr} also use the RL framework to learn a dynamic discounting scheme for PCFR$^+$, resulting in \textit{Dynamic PCFR$^+$ (DPCFR$^+$)}. PCFR$^+$ and DPCFR$^+$ are the prior practical SoTA for non-poker games, while in poker games, DCFR and DDCFR serve as the prior practical SoTA.

\section{Hyperparameter Schedules (HSs)}

\begin{figure*}[t]
    \centering
    \subfloat[Evolution of the weight $(t/(t+1))^\gamma$ applied to the contribution to the average strategy.\label{fig:gamma}]{
        \includegraphics[width=0.365\textwidth]{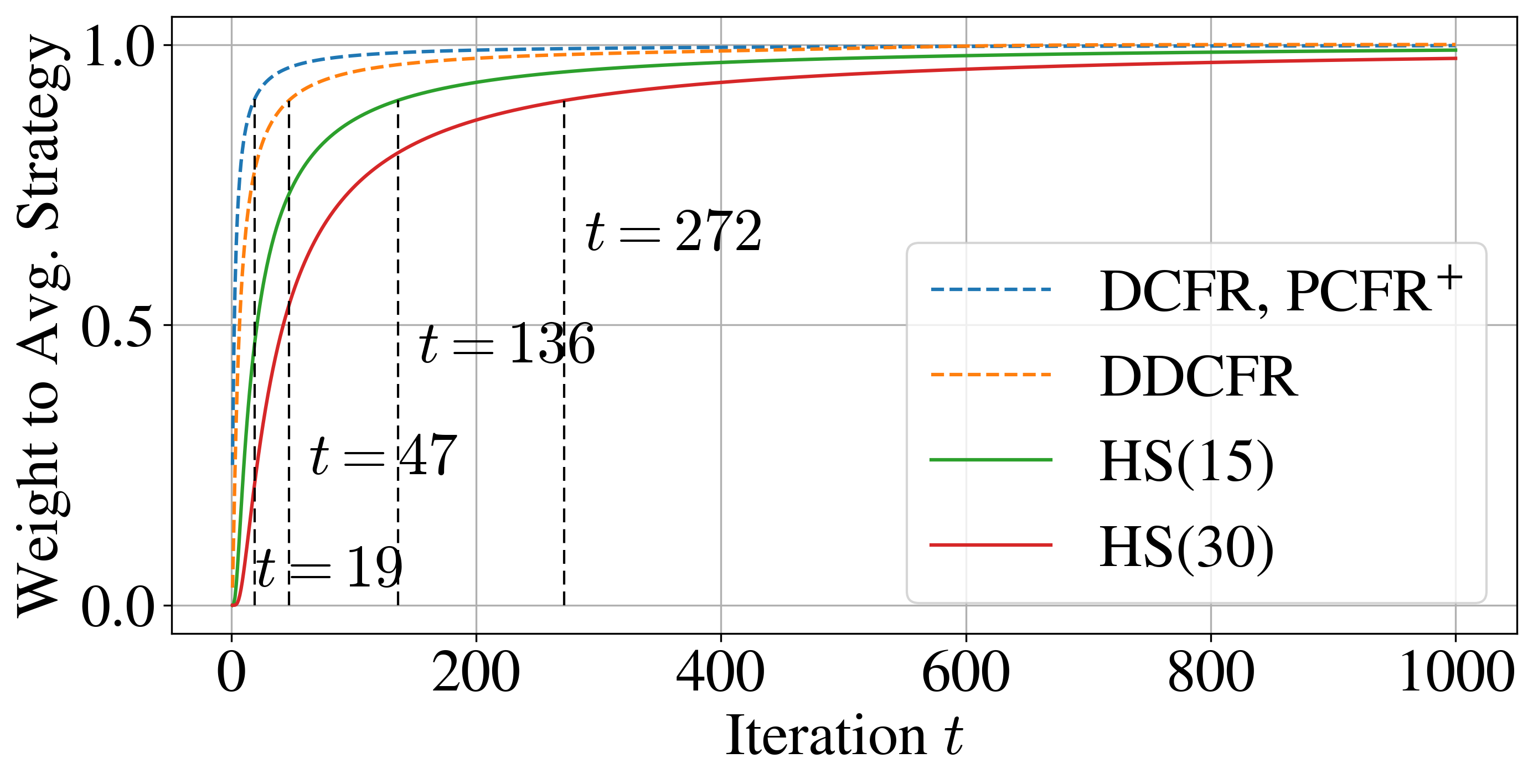}
    }\hfill
    \subfloat[Effect of $\gamma$ on the end exploitability of HS-DCFR at iteration 1,000.\label{fig:initial_gamma}]{
        \includegraphics[width=0.255\textwidth]{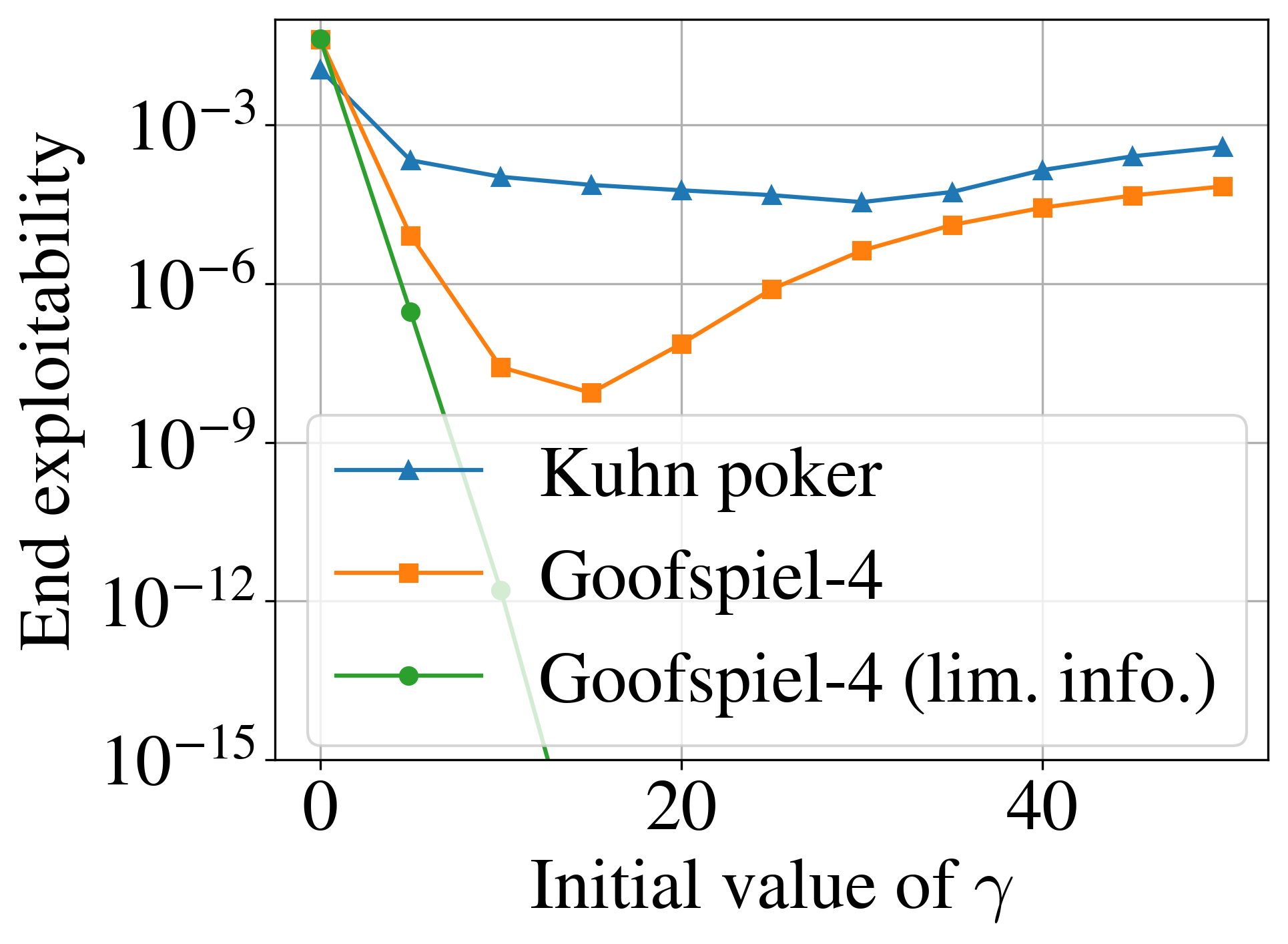}
    }\hfill
    \subfloat[Proposed HSs with a notably large $\gamma$ for 1,000 iterations.\label{fig:scheme}]{
        \includegraphics[width=0.25\textwidth]{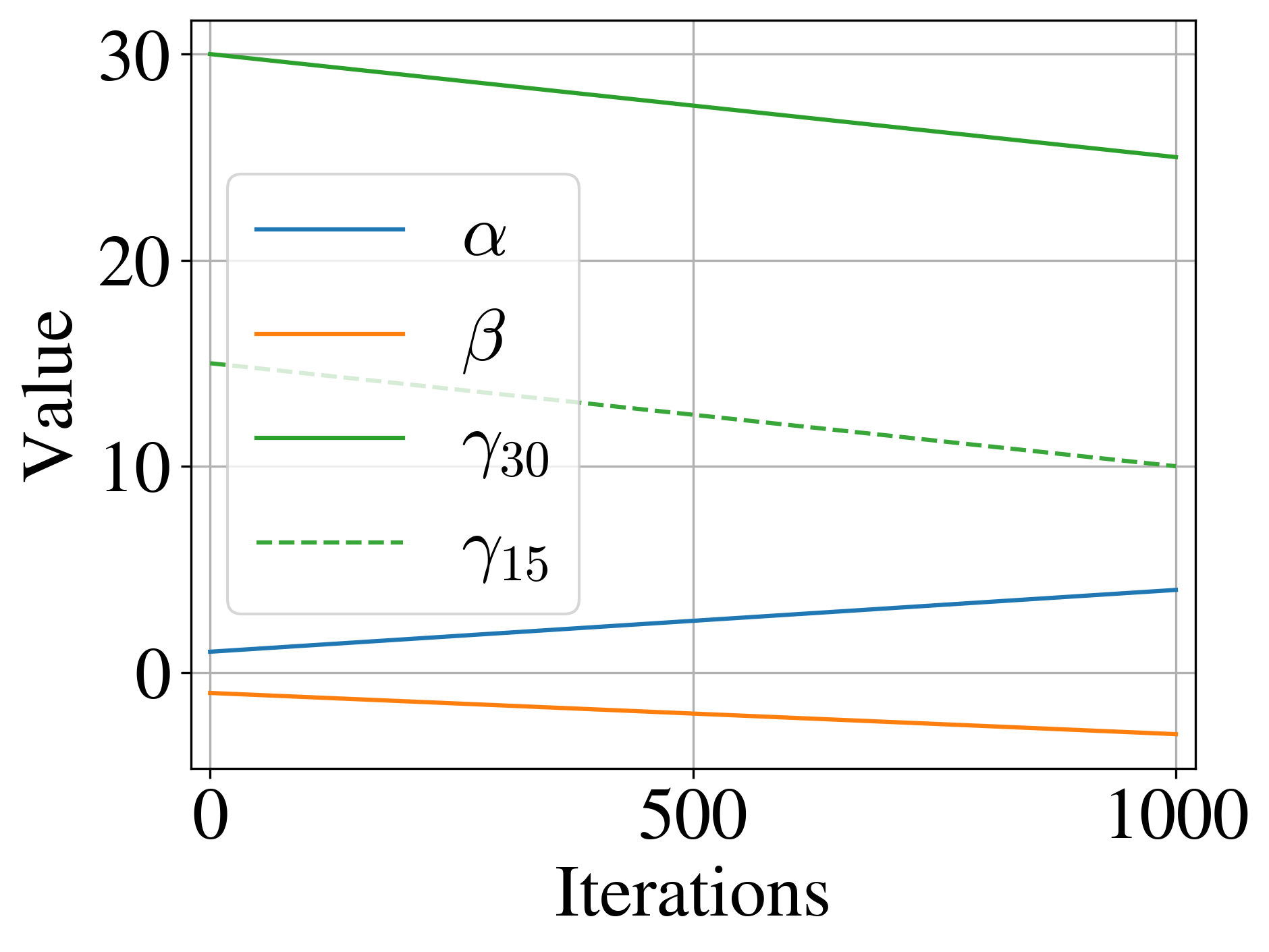}
    }
    \caption{Motivation and design of Hyperparameter Schedules (HSs).}
    \label{fig:test}
    \vspace{-1mm}
\end{figure*}

Many discounted CFR variants, such as CFR$^+$, DCFR, and PCFR$^+$, have demonstrated remarkable performance. However, they rely on a fixed discounting scheme, using the same hyperparameters across all iterations. For example, the authors of DCFR empirically determined that the best DCFR hyperparameters $(\alpha, \beta, \gamma)$ are $(1.5, 0, 2)$. Similarly, PCFR$^+$ fixes its discounting scheme by setting $\gamma$ to $2$ across all iterations.
In contrast, DDCFR proposes dynamically adjusting hyperparameters using an RL-based framework that trains an agent to determine a dynamic discounting scheme. Specifically, the DDCFR framework learns three HSs that vary each parameter non-linearly within the range $[-5, 5]$.

Although the motivation for dynamically changing hyperparameters during the run is valid, DDCFR incurs an overhead of additional compute and time required for feature calculations, policy training, and network inference for real-time computation of discounting weights. In particular, training the RL agent in their work took \textit{24 hours on 200 CPU cores}. 
Although the training time can be amortized if the trained discounting policy is reused across many games, when the characteristics of the games change over time or when new games with different dynamics are introduced, the pre-trained RL agent might not generalize well to the changes and necessitate retraining. 
Moreover, when using the DDCFR framework, \textit{multiprocessing} is required to run feature calculations concurrently to avoid increasing wall-clock time. In computationally constrained environments, these additional computational costs are inevitable. Beyond the computational overhead, the RL-based approach also introduces randomness inherent to the training process, so the performance of the learned discounting scheme can only be characterized within a confidence interval.

To leverage the benefits of a dynamic discounting scheme without the complexity and computational overhead of training and running an RL agent, we introduce the concept of \textit{Hyperparameter Schedules (HSs)}---discounting schemes that control how hyperparameters change across iterations---and investigate whether this lightweight, training-free approach can match or even surpass prior SoTA performance. 

\subsection{Motivation Behind HS$_{\gamma}$}

Although the aforementioned methods achieve strong practical performance, their discounting schemes are not sufficiently aggressive during the early stages of iterations. As a result, they assign substantial and persistent weight to the rough strategies produced in the initial iterations. To quantify this effect, Figure~\ref{fig:gamma} illustrates the evolution of the weight \((t/(t+1))^\gamma\), which determines the contribution of each iteration to the average strategy across different algorithms. As shown, all discounted CFR variants exhibit an inherent transition: early-stage updates contribute less to the average strategy, while later-stage updates have a greater influence (i.e., higher weight). However, there are notable differences in the \textit{rate} of transition. With a fixed \(\gamma=2\), both DCFR and PCFR\(^+\) rapidly approach a 0.9 weight within just 19 iterations. DDCFR, which employs a dynamic schedule for \(\gamma\), reaches the same threshold by 47 iterations---still within only 5\% of total training steps. Thus, the applied weight in prior methods converges to near 1 extremely early, meaning that, albeit discounted, the influence of early iterations on the average strategy is still strong.

Our key motivation is the need to delay the convergence of the weight toward 1 in order to \textit{aggressively} suppress the influence of early iterations and \textit{gradually} increase the contribution of later updates as CFR progresses. 
To counteract the impact of unrefined strategies from early iterations, we design our HSs for $\gamma$ to start at a very high value (e.g., 30). This forces the discount factor to stay well below 1 for hundreds of iterations, imposing a short memory that effectively suppresses the influence of early updates on the cumulative average strategy. As training progresses, the schedule gradually decreases $\gamma$, which increases the discount factor and extends the memory horizon, enabling the algorithm to incorporate the more stable and reliable strategies learned in later iterations.
This principled transition from aggressively downweighting early iterations to increasingly trusting late-stage updates is a key driver of the accelerated convergence observed in our experiments (Section~\ref{sec:exp}), and our extensive ablation studies empirically validate this principle.

\subsection{Identifying Effective HSs}
\label{sec:id_hs}

While our theoretical intuition suggests using a relatively large $\gamma$, there are additional intricacies in determining exactly how each schedule should be implemented. In this section, we describe the design choices and implementation details behind our scheduling strategies.
We refer to the combination of HS with DCFR as \textit{Hyperparameter Schedule-powered DCFR (HS-DCFR)}, which uses three HSs---one for each of $\alpha$, $\beta$, and $\gamma$. Constant schemes originally used in DCFR are special cases of HSs. For example, the empirically best scheme from the DCFR paper can be written as $\text{HS}_{\alpha} = 1.5$, $\text{HS}_{\beta} = 0$, and $\text{HS}_{\gamma} = 2$. We also combine HS with the prior algorithm that is SoTA in non-poker domains, PCFR$^+$. We call that combination  \textit{Hyperparameter Schedule-powered PCFR$^+$ (HS-PCFR$^+$)}. As discussed above, PCFR$^+$ uses one fixed HS for $\gamma$, namely $\text{HS}_{\gamma} = 2$. 

To identify effective HSs for HS-DCFR and HS-PCFR$^+$, we first experimented with large initial values of $\gamma$ ranging from $10$ to $50$ and found that HS-DCFR consistently outperformed DCFR across all tested games (see Section~\ref{sec:exp_setup} for game descriptions). In games such as Goofspiel-3, Goofspiel-4 (lim. info.), and Big Leduc poker, performance improves (i.e., achieving lower end exploitability) with increasing $\gamma$. However, for games such as Kuhn poker, Battleship-2, Battleship-3, other Goofspiel variants, and the Liar's dice family, performance initially improves and then degrades as $\gamma$ becomes too large. Figure~\ref{fig:initial_gamma} illustrates the effect of the initial value of $\gamma$ in HS on the end exploitability of HS-DCFR at 1,000 iterations. In Kuhn poker, initializing $\gamma$ at $30$ yields the best result, whereas in Goofspiel-4, $\gamma = 15$ is the preferred choice. Notably, for Goofspiel-4 (lim. info.), larger $\gamma$ values consistently lead to better performance, with $\gamma = 30$ already offering significant improvements. 

Therefore, we present two highly performant sets of HSs with a notably large $\gamma$ starting at $30$ or $15$, representing two distinct transition rates. As shown in Figure~\ref{fig:gamma}, HS(15) reaches a weight of 0.9 after 136 iterations, while HS(30) reaches a weight of 0.9 after 272 iterations---approximately one-third of the total iterations.
For DCFR, we also need to design HSs for $\alpha$ and $\beta$. However, our empirical findings indicate that the convergence speed of DCFR toward a Nash equilibrium is significantly more sensitive to the schedule of $\gamma$ than to those of $\alpha$ and $\beta$ (see Section~\ref{sec:ablt} and Appendix~\ref{apdx:ablt}). Therefore, we take inspiration from the parameter ranges used in DDCFR.
We denote the two HS-DCFR variants as HS-DCFR(30) that uses ($\text{HS}_\alpha$, $\text{HS}_\beta$, $\text{HS}_{\gamma{30}}$) and HS-DCFR(15) that uses ($\text{HS}_\alpha$, $\text{HS}_\beta$, $\text{HS}_{\gamma{15}}$). Figure~\ref{fig:scheme} illustrates each of the HSs and Equation~\ref{eq:hs} shows the exact formula, where $t$ is the current iteration and $n$ is the total number of iterations: 
\begin{equation}
\label{eq:hs}
\begin{aligned}
    & \text{HS}_\alpha: \alpha = 1+\frac{3}{n}\ t,              && \text{HS}_\beta: \beta = -1-\frac{2}{n}\ t, \\ 
    & \text{HS}_{\gamma{30}}: \gamma_{30} = 30-\frac{5}{n}\ t,  && \text{HS}_{\gamma{15}}: \gamma_{15} = 15-\frac{5}{n}\ t. \\
\end{aligned}
\end{equation}
For example, to implement HS-DCFR(30), one simply plugs the formulae from Equation~\ref{eq:hs} into Equations~\ref{eq:dcfr_alpha_beta} and~\ref{eq:dcfr_gamma} for $\alpha$, $\beta$, and $\gamma$.
Additionally, we apply the same two HSs for adjusting $\gamma$ to PCFR$^+$, yielding HS-PCFR$^+$(30) that uses $\text{HS}_{\gamma{30}}$ and HS-PCFR$^+$(15) that uses $\text{HS}_{\gamma{15}}$. Given that $\gamma$ is the only adjustable hyperparameter in PCFR$^+$, HSs for $\alpha$ and $\beta$ are not used in this setting.

We do not claim that the proposed HSs are optimal for any specific game, as there are infinitely many possible scheduling curves and our schedules are not tuned to individual games. 
Rather, we present two example schedules to demonstrate that it is possible to design training-free, easy-to-implement, and computationally efficient approaches (requiring only minimal overhead of accessing pre-determined schedule values at runtime) that significantly accelerate CFR convergence across a diverse set of games. We provide the exact formulae in Equation~\ref{eq:hs} so that readers can reproduce and adopt them for faster game solving, as the required code modifications are minimal (fewer than 15 lines). Others may further build on our schedules, using additional time and computational resources to develop more fine-tuned variants for each game or game class.

\subsection{Theoretical Analysis of Convergence Rates}
\label{sec:theorems}

We prove the worst-case convergence rate of HS-DCFR given that the hyperparameters $\alpha$, $\beta$, and $\gamma$ fall within certain ranges. We also establish the worst-case convergence rate of HS-PCFR$^+$ given that the hyperparameter $\gamma$ falls within a certain range. Both proofs adopt a common simplification that is used in prior work~\cite{tammelin2014solving,brown2019solving,ddcfr}, namely assuming that both players update their regrets simultaneously on each iteration. Proofs are provided in Appendix~\ref{apdx:proof} and~\ref{apdx:proof_thm2}.
\begin{theorem}
\label{thm:hsdcfr}
Suppose $T$ iterations of HS-DCFR, with simultaneous updates, are played in a two-player zero-sum game, and $U$ is the upper bound of $\gamma$ across all iterations. If $\alpha \in [1, 5]$, $\beta \in [-5, 0]$, and $\gamma \in [0, U]$, the weighted average strategy profile is a \linebreak $(U+1) \Delta|\mathcal{I}|\left(\frac{8}{3} \sqrt{|\mathcal{A}|}+\frac{2}{\sqrt{T}}\right) / \sqrt{T}$-Nash equilibrium. 
\end{theorem}

\begin{theorem}
\label{thm:hspcfr}
Suppose $T$ iterations of HS-PCFR$^+$, with simultaneous updates, are played in a two-player zero-sum game, and $U$ is the upper bound of $\gamma$ across all iterations. If $\gamma \in [0, U]$, the weighted average strategy profile is a \linebreak $(U+1) |\mathcal{I}| O(1)/ \sqrt{T}$-Nash equilibrium. 
\end{theorem}

\section{Experiments}
\label{sec:exp}

\begin{figure*}[t]
\centering
    \subfloat{\includegraphics[width=0.438\columnwidth,trim={6 10 8 6},clip]{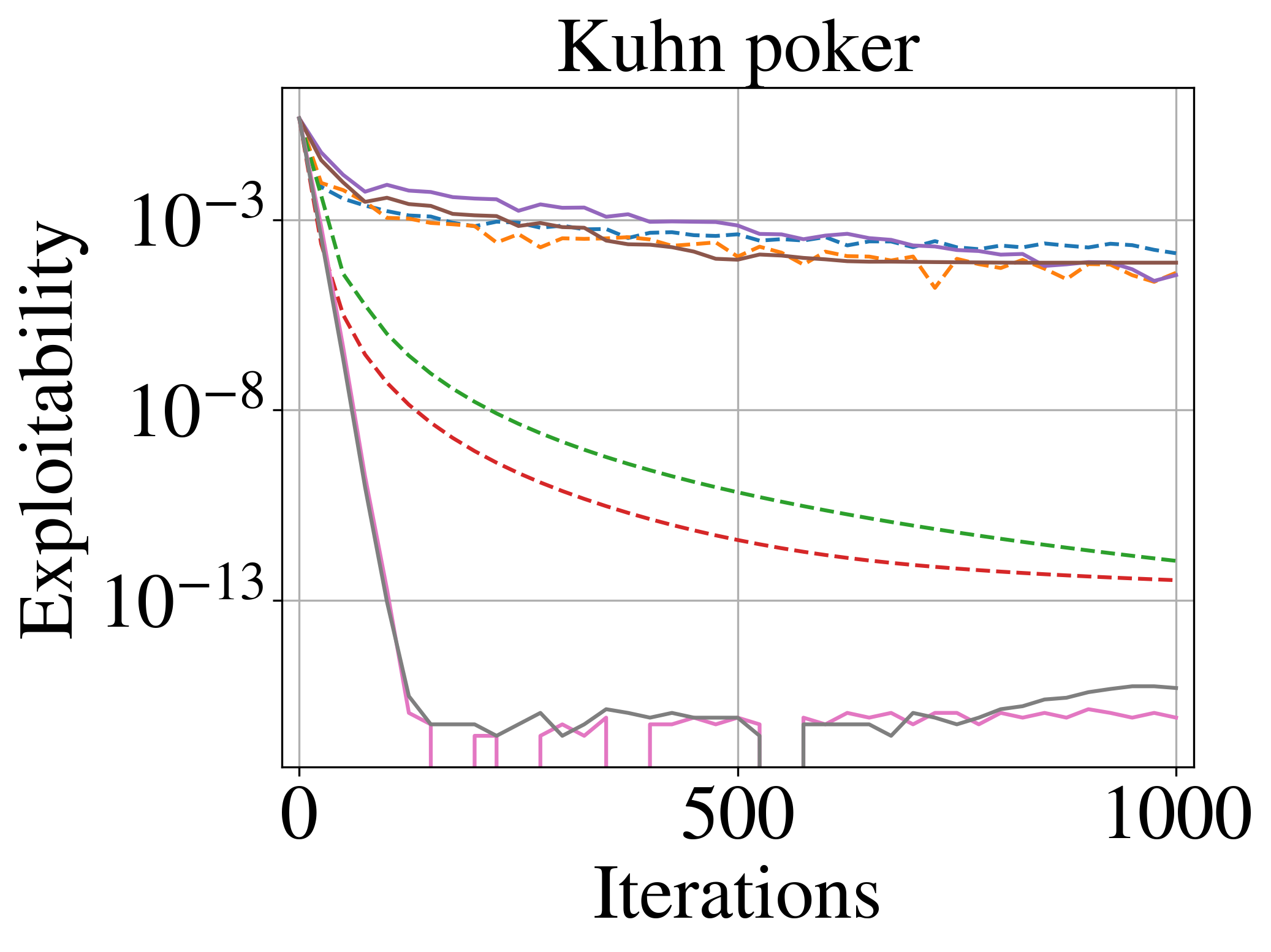}}
    \subfloat{\includegraphics[width=\figscale\columnwidth,trim={6 10 8 6},clip]{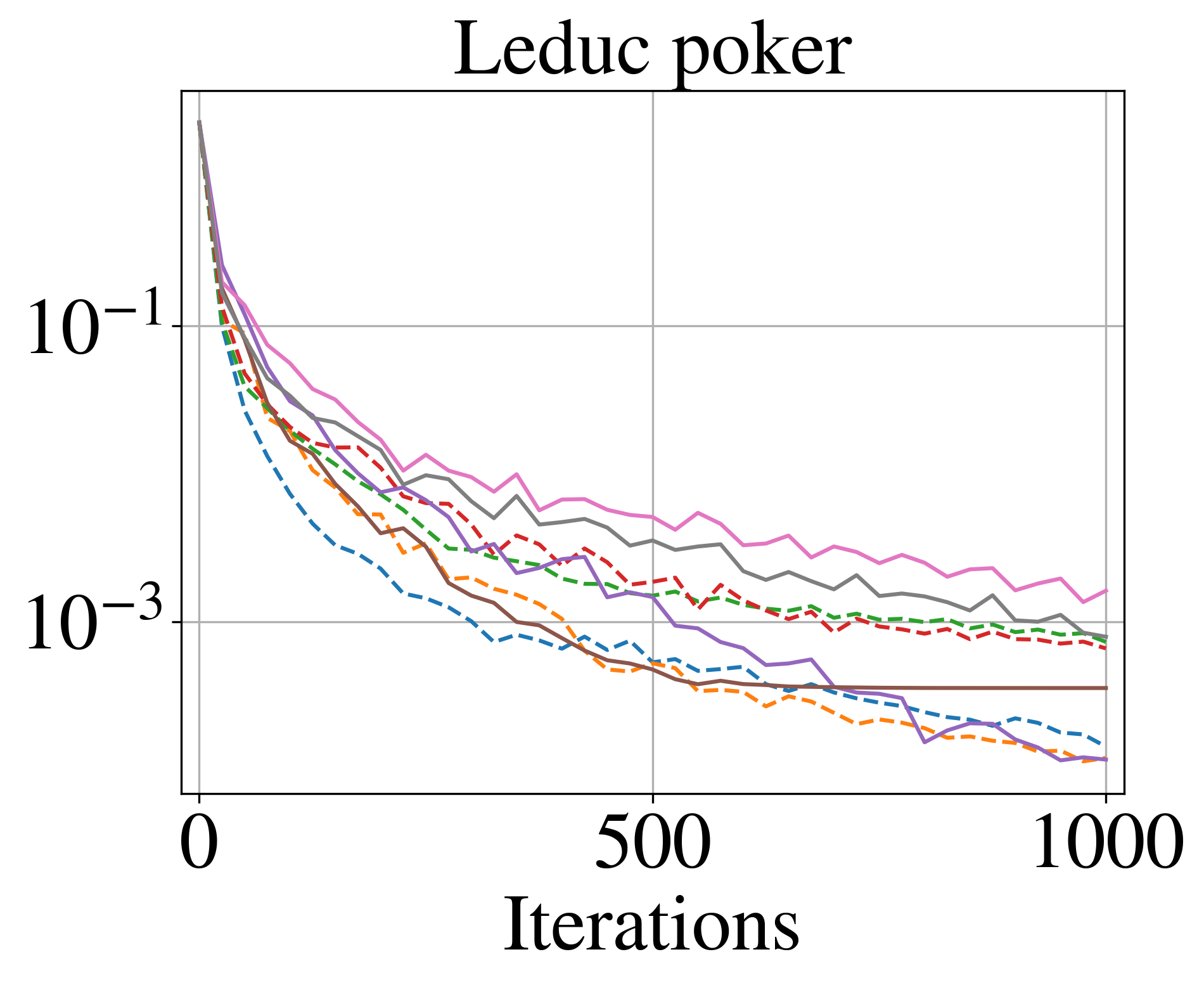}}
    \subfloat{\includegraphics[width=0.41\columnwidth,trim={6 10 8 6},clip]{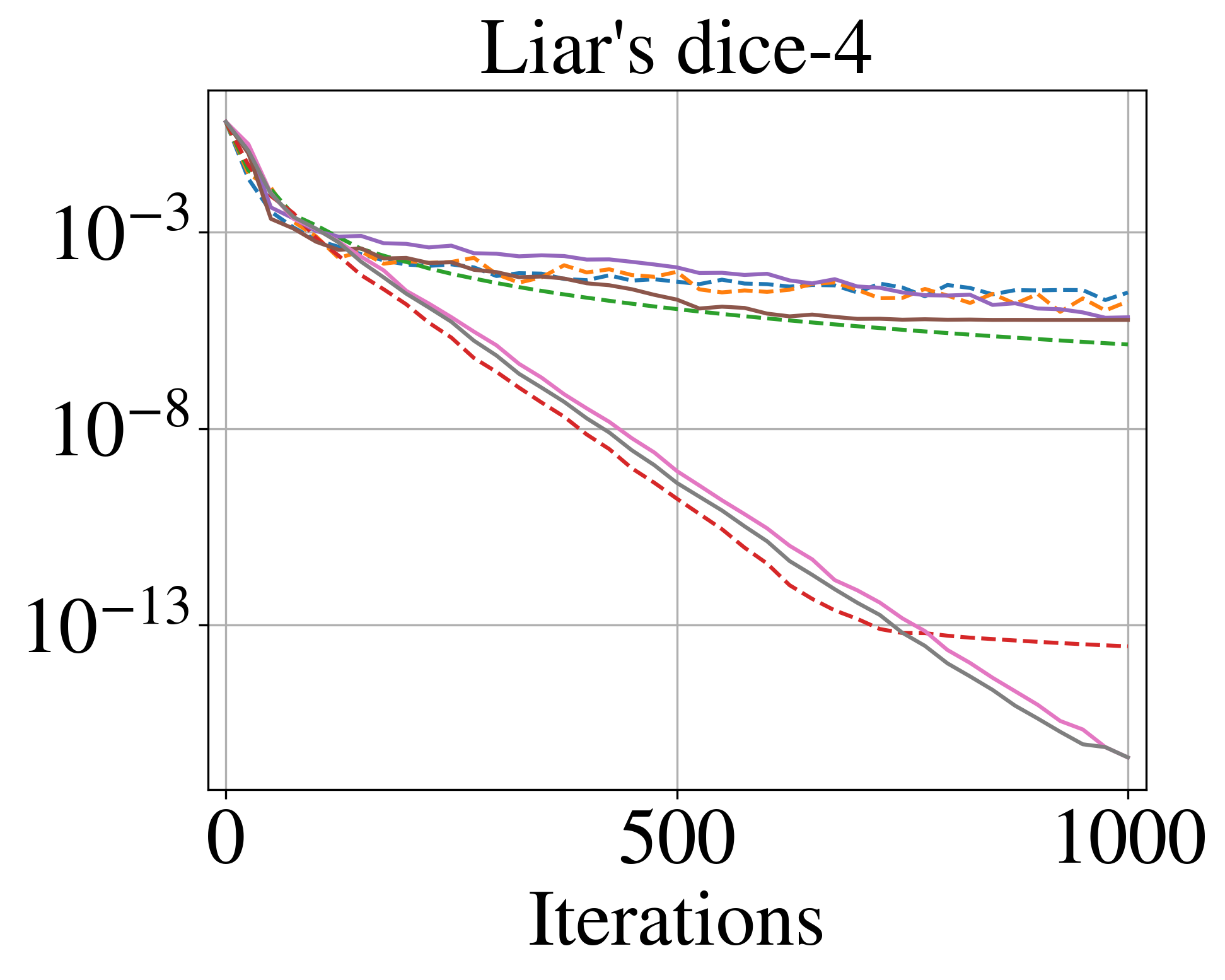}} 
    \subfloat{\includegraphics[width=\figscale\columnwidth,trim={6 10 8 6},clip]{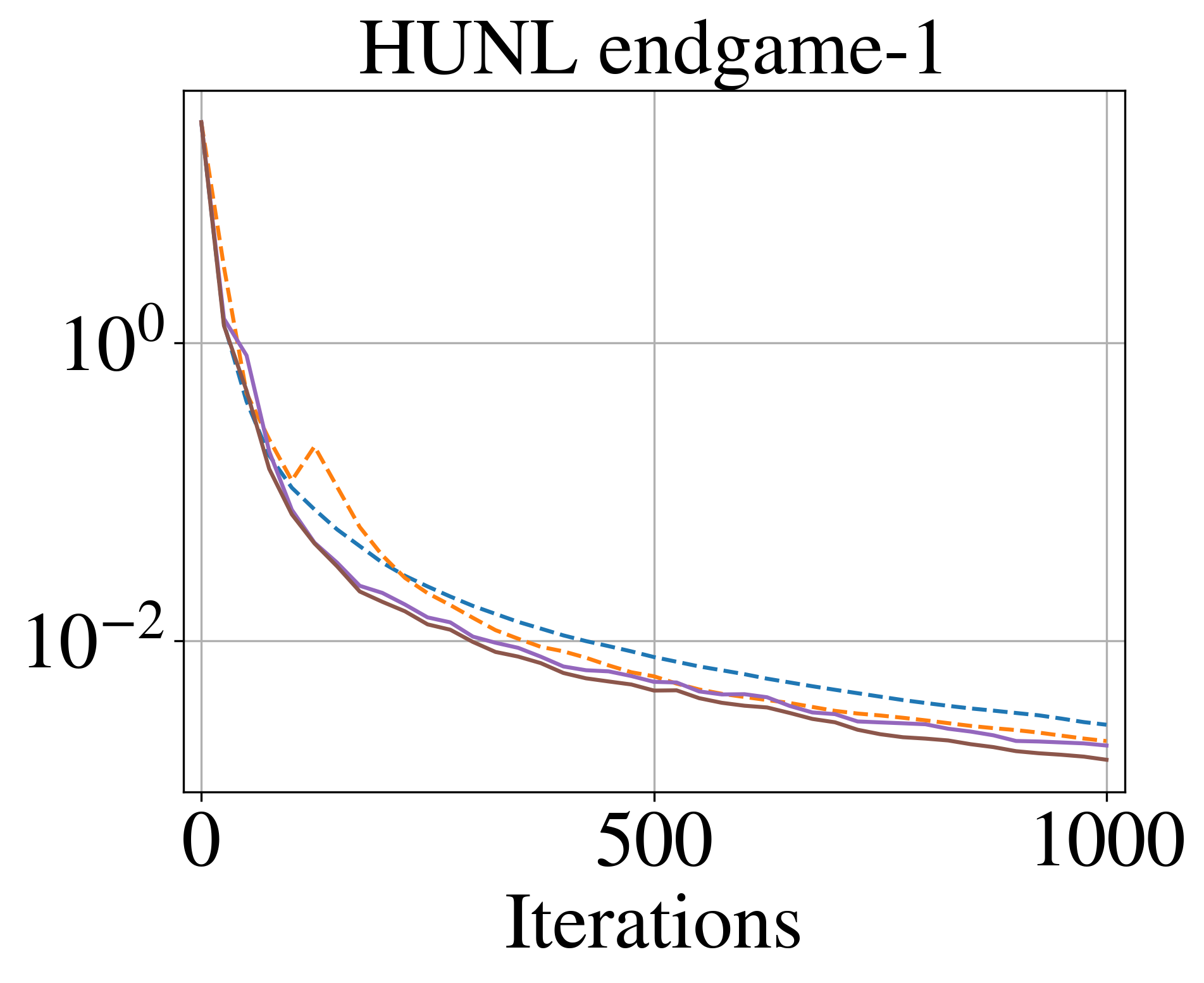}}
    \subfloat{\includegraphics[width=\figscale\columnwidth,trim={6 10 8 6},clip]{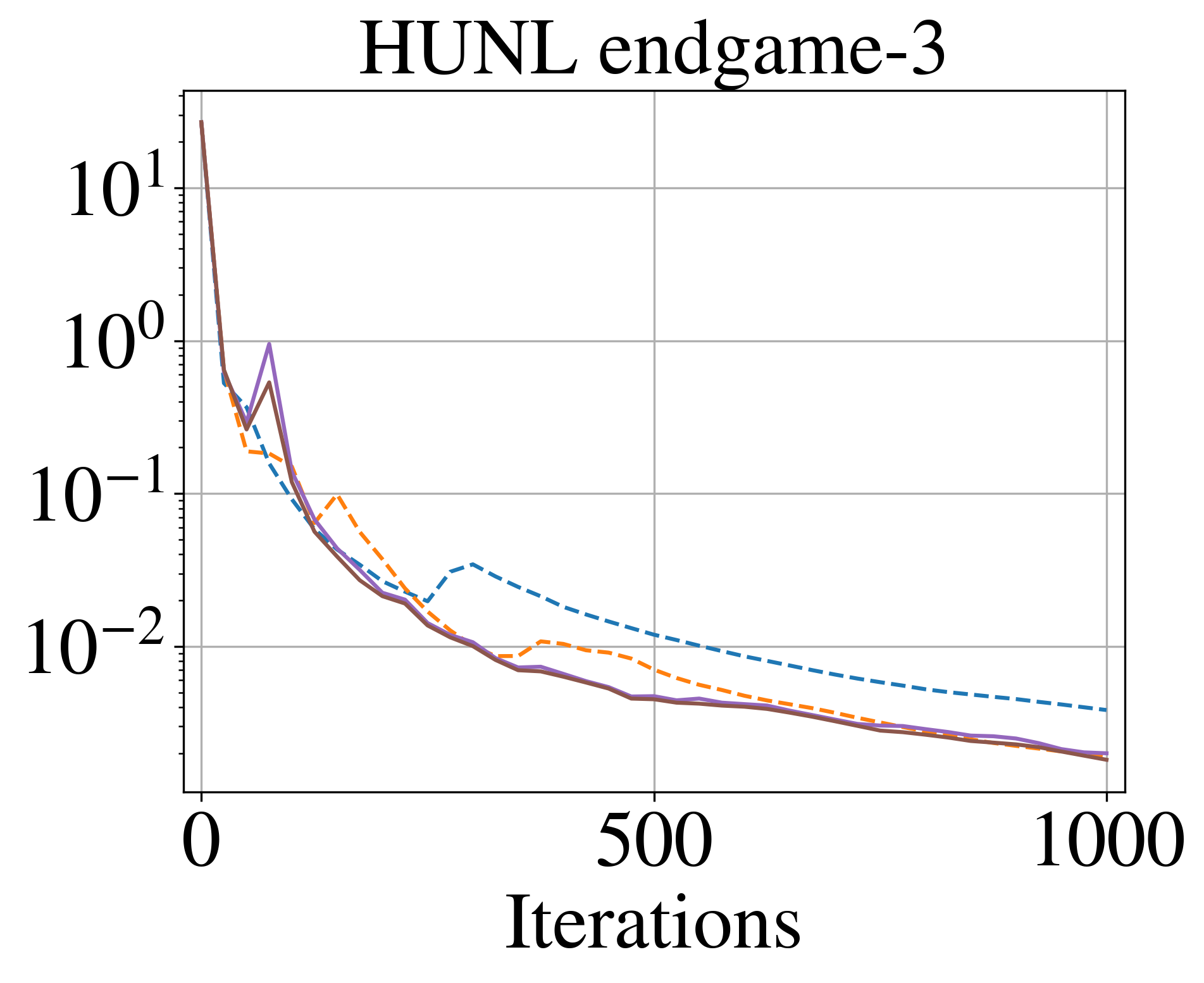}} \\ 
    \subfloat{\includegraphics[width=0.438\columnwidth,trim={6 10 8 6},clip]{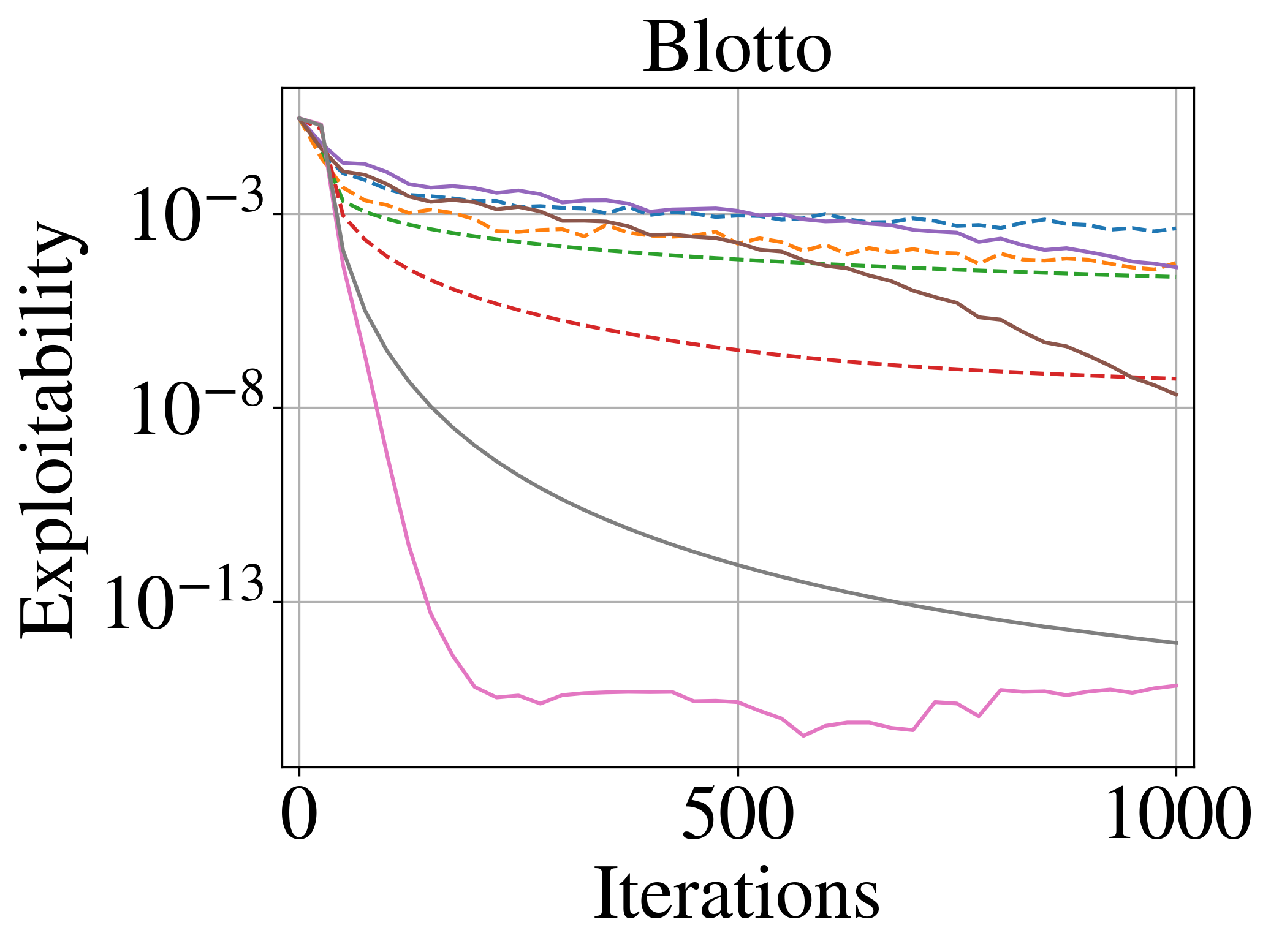}}
    \subfloat{\includegraphics[width=\figscale\columnwidth,trim={6 10 8 6},clip]{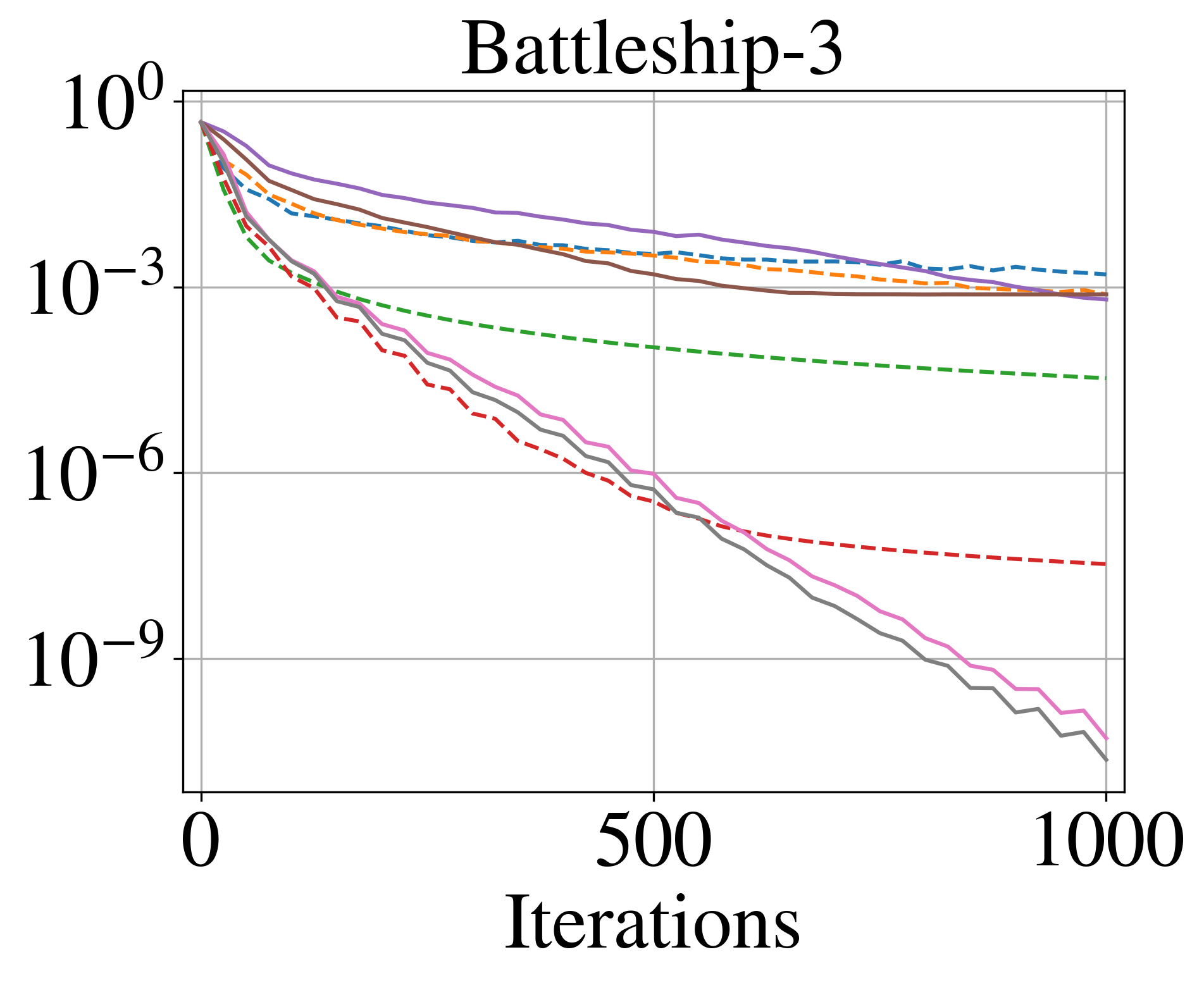}}
    \subfloat{\includegraphics[width=0.41\columnwidth,trim={6 10 8 6},clip]{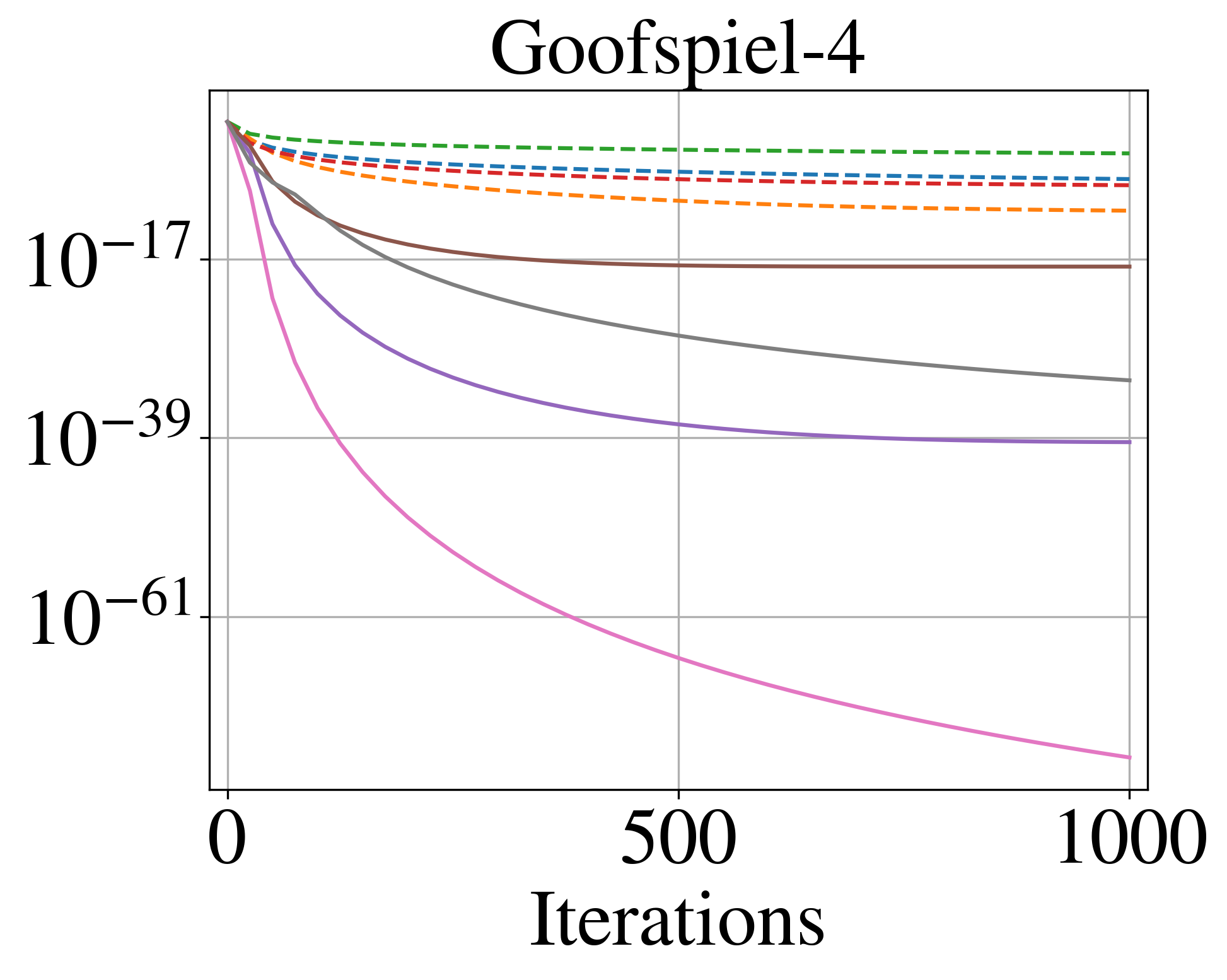}} 
    \subfloat{\includegraphics[width=0.41\columnwidth,trim={6 10 8 6},clip]{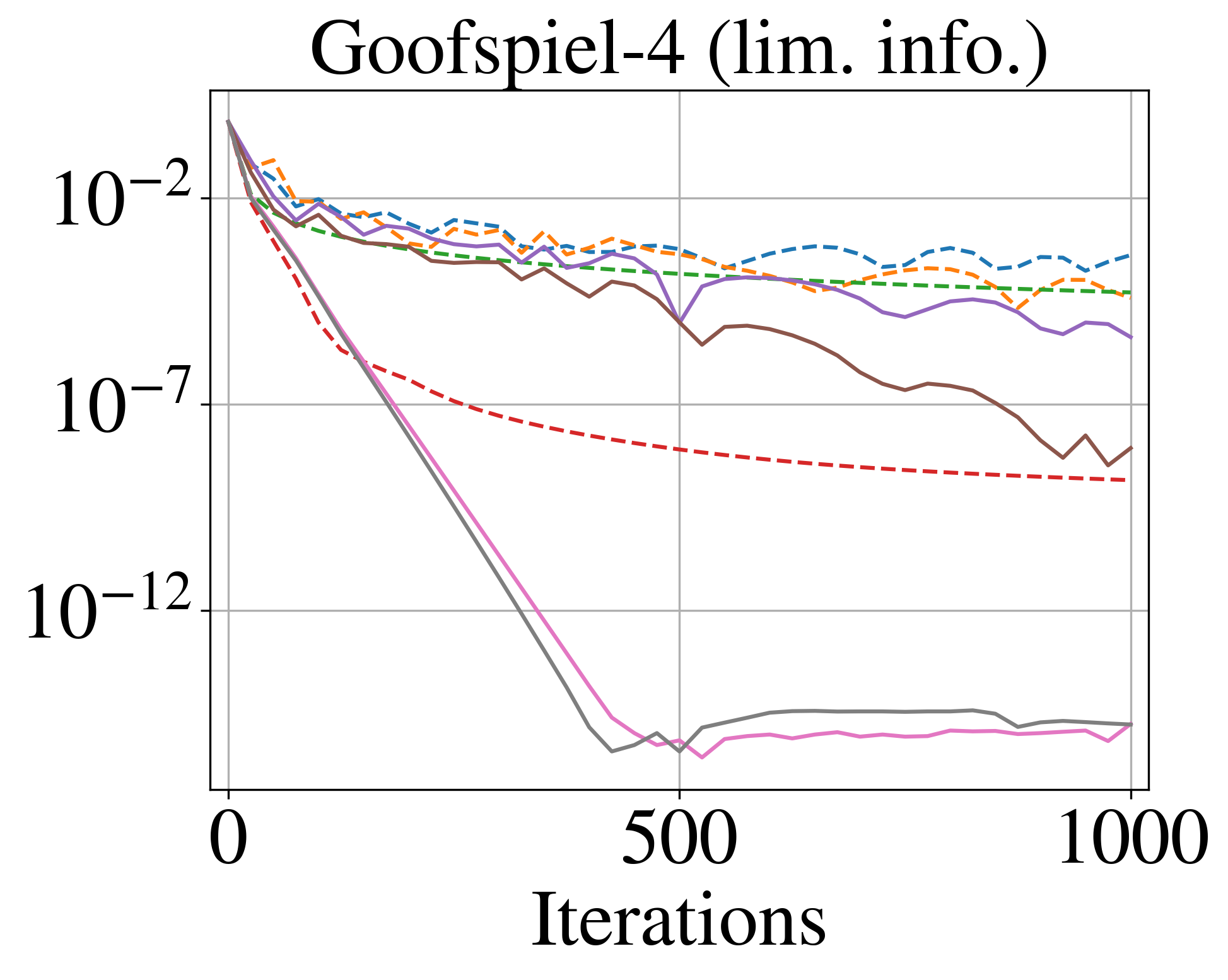}} 
    \subfloat{\includegraphics[width=0.41\columnwidth,trim={6 10 8 6},clip]{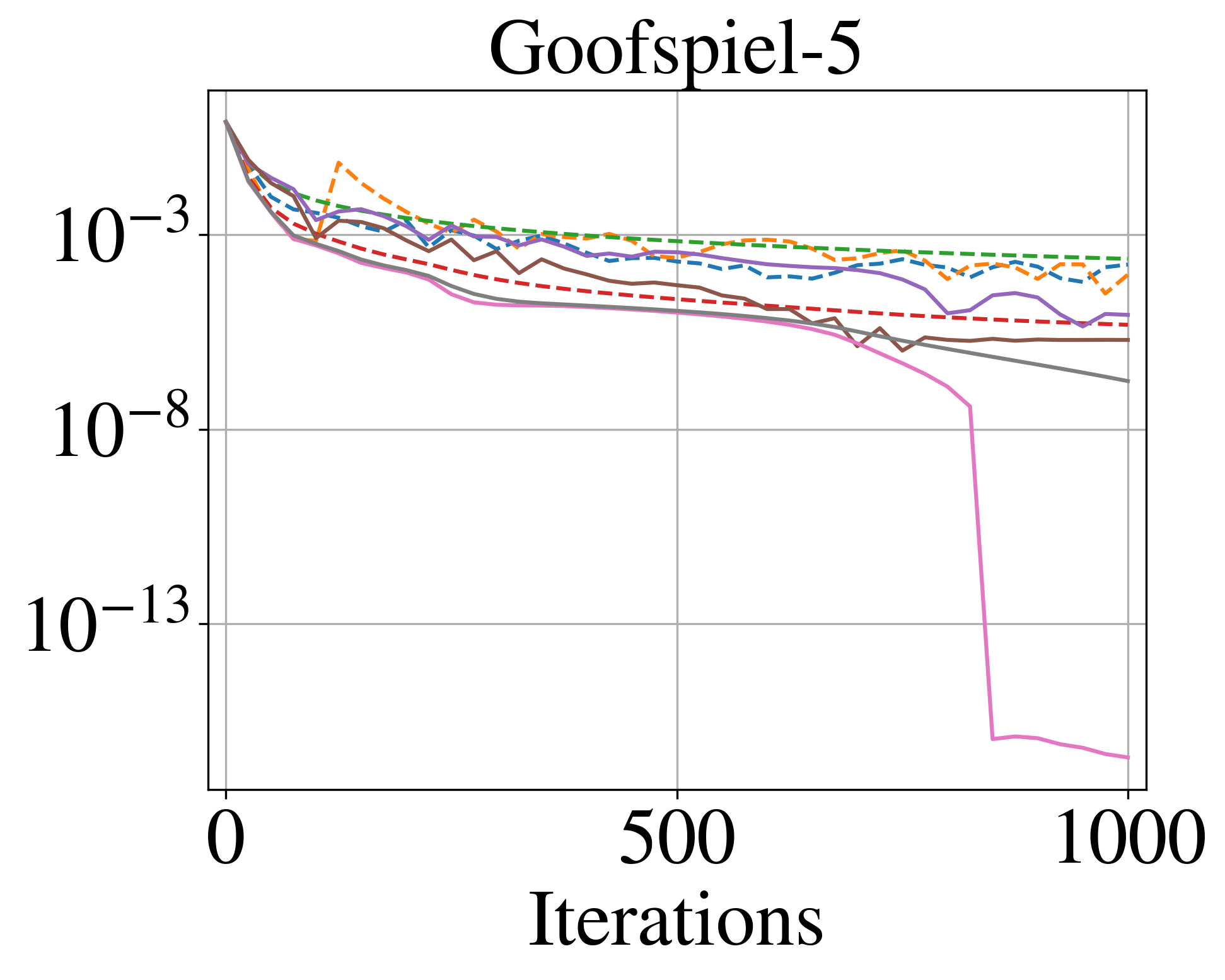}} \\
    \subfloat{\includegraphics[width=0.8\linewidth,trim={0 125 0 125},clip]{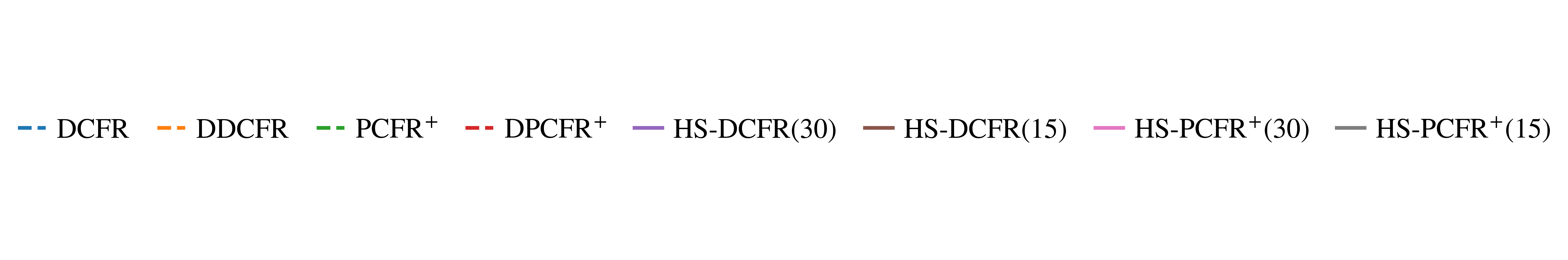}}
    \caption{Performance of HS-powered algorithms and prior SoTA algorithms on extensive-form and normal-form games.}
    \label{fig:results}
    \vspace{-1mm}
\end{figure*}

In this section, we describe our experimental setup, followed by performance comparisons of HS-DCFR and HS-PCFR$^+$ against prior state-of-the-art algorithms.

\subsection{Experimental Setup}
\label{sec:exp_setup}

For evaluation, we use nine extensive-form IIGs and one normal-form IIG that are widely used as benchmarks in computational equilibrium finding. 
These games include Kuhn poker, Leduc poker, Liar's dice-4, Battleship-3, Goofspiel-4 (with and without limited information), Blotto (a normal-form IIG), Goofspiel-5 (without limited information), \textit{heads-up no-limit Texas hold 'em (HUNL)} endgame-1, and HUNL endgame-3. Additional results on seven more IIGs are provided in Appendix~\ref{apdx:results}.
This diverse set encompasses games of varying complexity, size, and structure, enabling a comprehensive demonstration of the effectiveness of HS-powered algorithms. 

Here, we provide a brief overview of the games and their variants. 
Detailed descriptions and game size measurements are provided in Appendix~\ref{apdx:game_desc}.
Kuhn poker~\cite{kuhn1950simplified} is a classic three-card toy poker game with a single betting round.
Battleship-$x$~\cite{farina2019correlation} is a simplified traditional board game where players secretly position a single ship on their $2 \times x$ grid and alternate taking shots at each other's ship.
Goofspiel-$x$~\cite{ross1971goofspiel} is a bidding card game, where each player has $x$ cards and, by placing sealed bids, attempts to score the highest number of points over $x$ rounds. 
Goofspiel-$x$ (lim. info.)~\cite{lanctot2009monte} is the limited-information variant where the players withhold their card information during each turn. 
Leduc poker~\cite{southey2005bayes} is an extended version of Kuhn poker with six cards in the deck divided into two suits. There are two betting rounds and a maximum of two raises per round. 
Big Leduc poker expands Leduc poker to a deck of 24 cards divided into two suits, allowing a maximum of six raises per round. 
Liar's dice-$x$~\cite{lisy2015online} is a game where each player rolls an $x$-sided die, and the players take turns bidding on the outcome. 
HUNL endgame-$x$ represents an endgame scenario of heads-up no-limit Texas hold 'em. 
Blotto~\cite{golman2009general} is a resource allocation game where each player allocates finite resources across multiple battlefields.

We implemented four HS-powered algorithms and to the best of our ability reproduced four baselines using OpenSpiel v1.2~\cite{lanctot2019openspiel} and PokerRL v0.03~\cite{steinberger2019pokerrl}.
For HUNL endgames, we plot only the performance of the DCFR variants due to the complexity of reproducing PCFR$^+$ variants in large poker games, as their code is not open-sourced. Notably, as reported in the PCFR$^+$ paper, DCFR outperforms PCFR$^+$ on HUNL endgames.
Due to the simplicity of HS, we were able to run the experiments on a single Apple M2 CPU with 16 GB of RAM, with \mbox{macOS} Sequoia v15.5. We set the number of iterations to 1,000, which is sufficient to reach low exploitability and is a typical stopping time used in CFR practice. 
We used alternating updates in all of the algorithms; that is, in each iteration, both players perform an update one after the other. All algorithms are deterministic; therefore, each algorithm is run once and no confidence intervals are reported. 

\subsection{Experimental Results}

In this section, we present our experimental results for extensive-form games, demonstrating that our proposed algorithms constitute the new SoTA. 
In Figure~\ref{fig:results}, we compare four HS-powered algorithms (HS-DCFR(30), HS-DCFR(15), HS-PCFR$^+$(30), and HS-PCFR$^+$(15)) against the prior SoTA algorithms (DCFR, DDCFR, PCFR$^+$, and DPCFR$^+$). These results show that HS-based methods achieve strong performance across a diverse set of games. The new algorithms outperform the prior SoTA by many orders of magnitude on many games and are at least competitive on the remaining ones. 
To quantify the performance gap, we compute the ratio between the minimum exploitability ($\mathcal{E}$) achieved by DCFR, DDCFR, PCFR$^+$, or DPCFR$^+$ and that achieved by HS-PCFR$^+$(30). We then take the logarithm (base 10) of this ratio, defining the \textit{order of magnitude (OoM)} difference as $\log_{10}(\mathcal{E}_{\text{SoTA}} / \mathcal{E}_{\text{HS-PCFR$^{+}$(30)}})$, where $\mathcal{E}_{\text{SoTA}} = \min(\mathcal{E}_{\text{DCFR}}, \mathcal{E}_{\text{DDCFR}}, \mathcal{E}_{\text{PCFR$^{+}$}}, \mathcal{E}_{\text{DPCFR$^{+}$}})$. Across the ten test games, HS-PCFR$^+$(30) significantly outperforms the prior SoTA by an average of \textit{$12.5$ orders of magnitude}. In the following subsections, we discuss the performance of HS-DCFR and HS-PCFR$^+$ in more detail.

\begin{figure*}[t]
\centering
    \subfloat{\includegraphics[width=0.438\columnwidth,trim={6 10 8 6},clip]{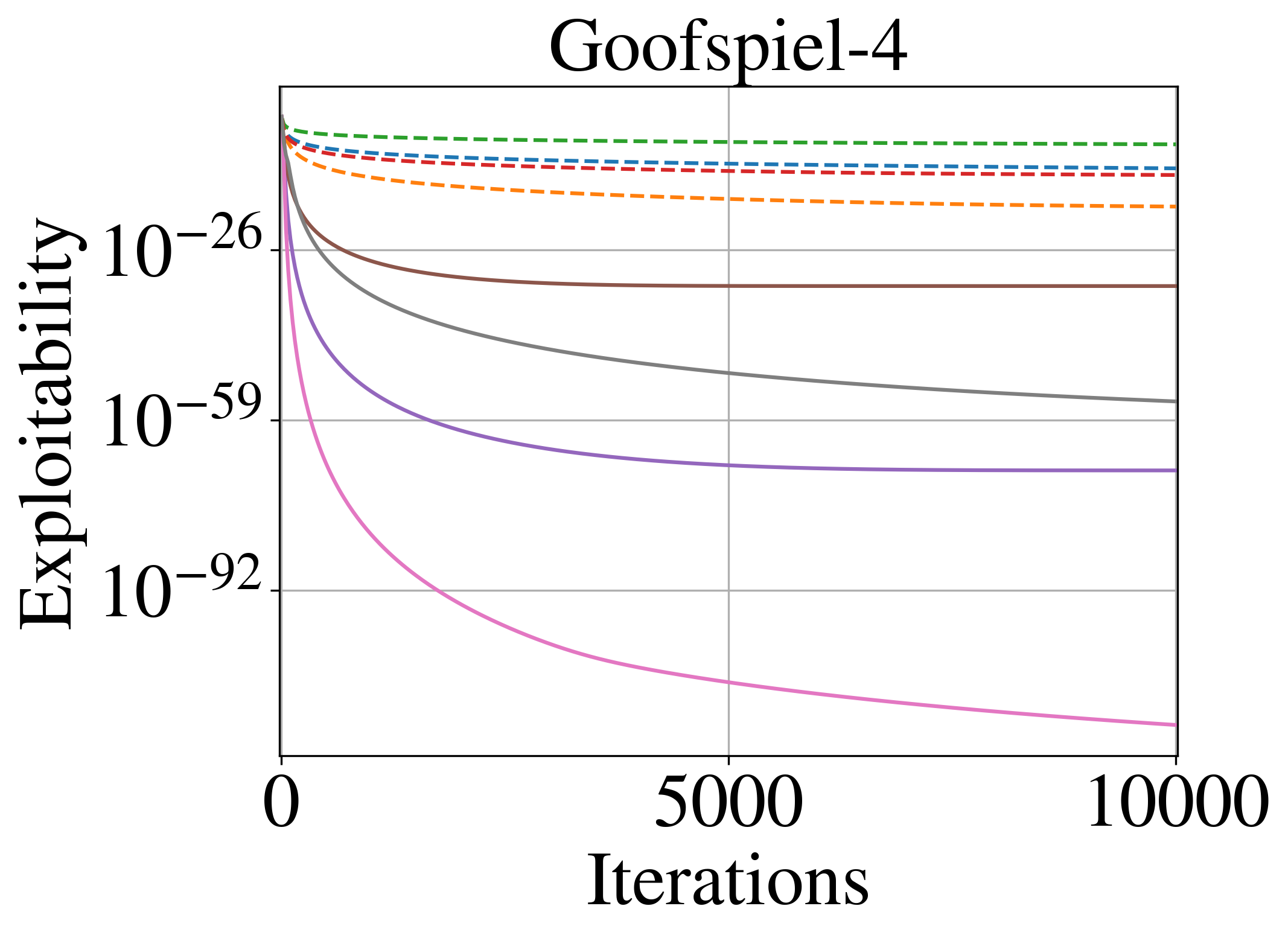}}
    \subfloat{\includegraphics[width=0.41\columnwidth,trim={6 10 8 6},clip]{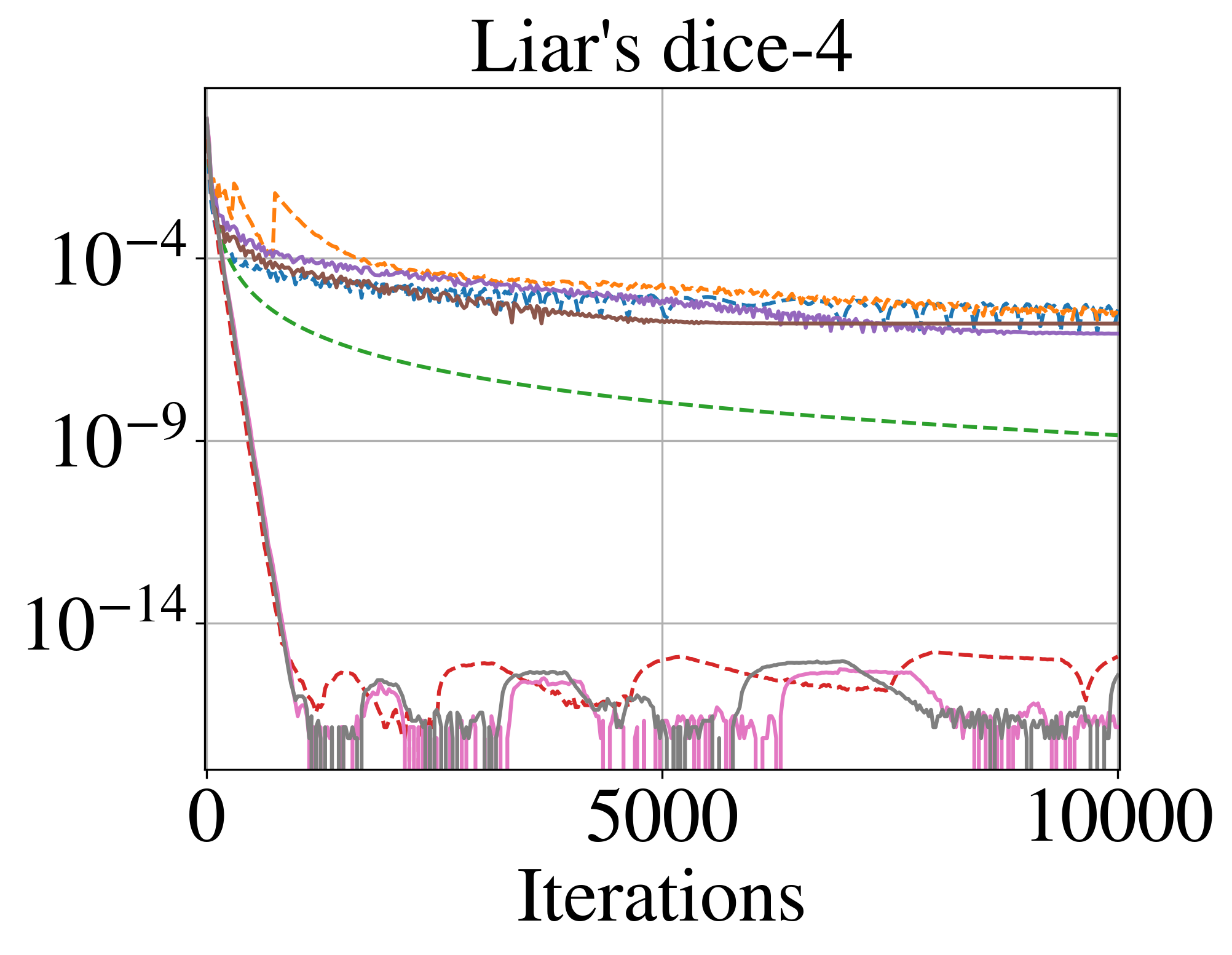}}
    \subfloat{\includegraphics[width=0.4\columnwidth,trim={6 10 8 6},clip]{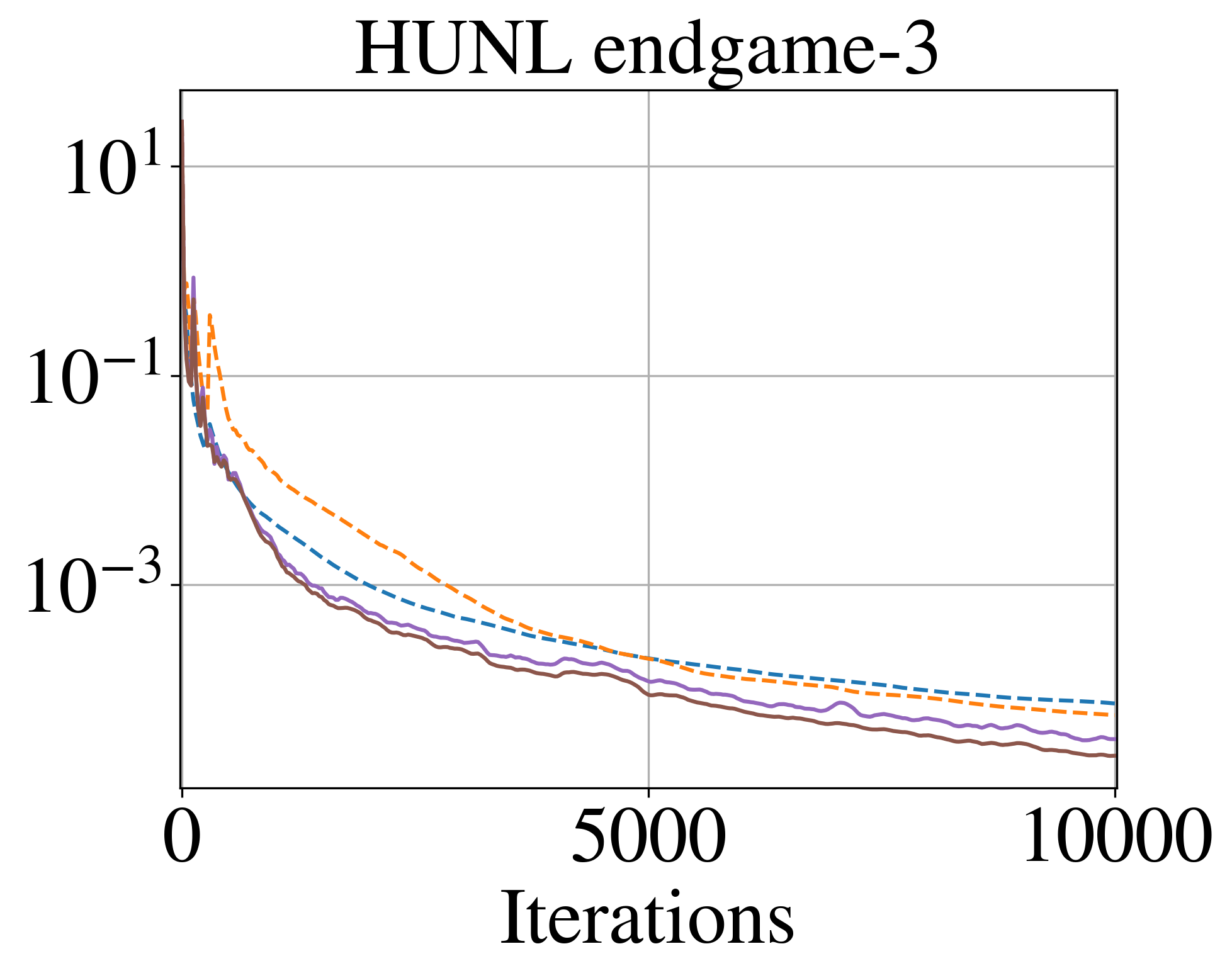}}
    \subfloat{\includegraphics[width=0.41\columnwidth,trim={6 10 8 6},clip]{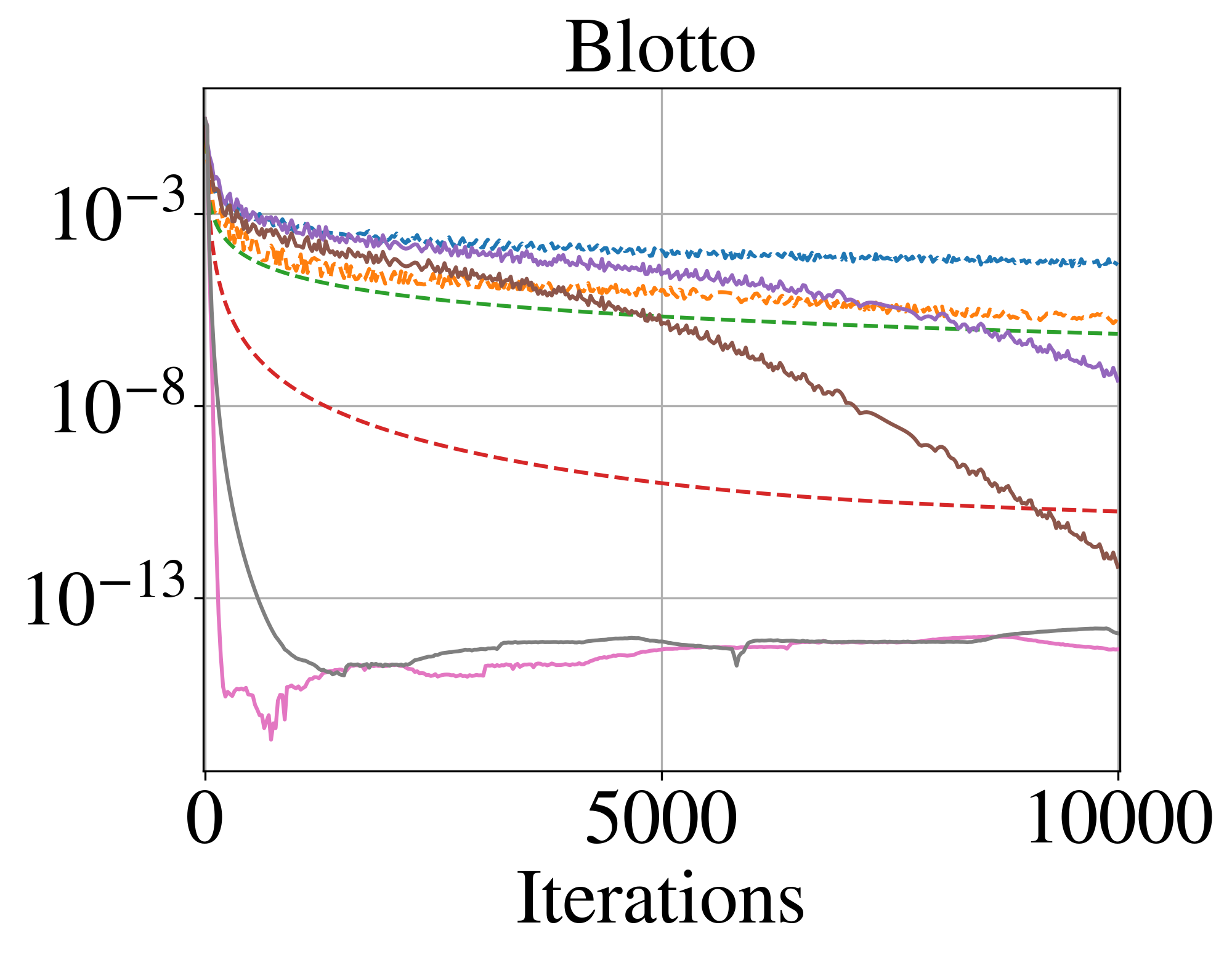}} \\
    \subfloat{\includegraphics[width=0.8\linewidth,trim={0 125 0 125},clip]{Plots/legend.png}}
    \caption{Performance of HS-powered algorithms and prior SoTA algorithms with an extended number of iterations.}
    \label{fig:games_10k}
    \vspace{-1mm}
\end{figure*}

\subsubsection{HS-DCFR}

Across the ten benchmark games, HS-DCFR(30) outperforms DDCFR by an average of $3.1$ OoM, while HS-DCFR(15) outperforms DDCFR by an average of $1.6$ OoM. In HUNL endgames, both HS-DCFR variants consistently achieve strong performance compared to the prior SoTA. These results indicate that simply integrating HS into the original DCFR algorithm yields superior performance compared to DDCFR, despite the latter incurring additional computational costs and complexity. Furthermore, whereas DDCFR employs an RL framework to generate an individual hyperparameter schedule for each game (i.e., game-specific tuning), our approach applies one HS across all games, either ($\text{HS}_\alpha$, $\text{HS}_\beta$, $\text{HS}_{\gamma{30}}$) or ($\text{HS}_\alpha$, $\text{HS}_\beta$, $\text{HS}_{\gamma{15}}$), without any game-specific or game-class-specific tuning. 

\subsubsection{HS-PCFR$^+$}

Across the ten games, HS-PCFR$^+$(30) emerges as the new SoTA, surpassing the prior SoTA by an average of $12.5$ OoM. HS-PCFR$^+$(15) also performs strongly, outperforming the prior SoTA by $5.4$ OoM on average. Notably, in games like Battleship-3 and Liar's dice-4, the exploitability of the HS-PCFR$^+$ variants decreases at an exponential rate without plateauing by 1,000 iterations. In Goofspiel-5, HS-PCFR$^+$(30) exhibits a substantial drop in exploitability around iteration 800, indicating the discovery of better strategies that traditional CFR variants fail to uncover. 
Although HS-PCFR$^+$(30) outperforms HS-DCFR(30) by an average of $12.4$ OoM across the benchmark games, HS-DCFR(30) is more than an OoM better than HS-PCFR$^+$(30) in Leduc, Big Leduc, and Liar's dice-6. 
This aligns with the observation in the PCFR$^+$ paper that PCFR$^+$ is less effective than DCFR in poker-like games (except for Kuhn poker, a tiny game). Therefore, we recommend using HS-DCFR(30) for large poker games and HS-PCFR$^+$(30) for the remaining games.

\subsubsection{Normal-Form Game}

We also test HS-powered algorithms on a common \textit{normal-form} zero-sum benchmark game, General Blotto~\cite{golman2009general}. In Blotto, each player allocates finite resources across multiple battlefields, with each battlefield awarded to the player allocating more resources; the final payoff is proportional to the number of battlefields won.
We use the setting with two players, three battlefields, and five resources per player. Remarkably, HS-PCFR$^+$(30) surpasses the prior SoTA by $7.9$ OoM, establishing the new SoTA on this normal-form zero-sum game by a large margin.

\subsubsection{Extended Iterations}

In Figure~\ref{fig:games_10k}, we increase the number of iterations to 10,000 when running all algorithms. HS-powered algorithms consistently outperform the prior SoTA by a significant margin. For HUNL endgame-3, we observe a larger performance gain compared to running the algorithms for only 1,000 iterations. In Liar's dice-4, the HS-PCFR$^+$ variants produce a jagged curve, possibly due to instability during updates when regret values are extremely small.

\subsubsection{Choice of $\gamma$}
\label{sec:choosing_gamma}

We propose two variants for each of HS-DCFR and HS-PCFR$^+$, using either $\text{HS}_{\gamma{30}}$ or $\text{HS}_{\gamma{15}}$. Our experiments show that the two schedules exhibit selective superiority: when $\text{HS}_{\gamma{15}}$ performs better, the gains are modest, whereas when $\text{HS}_{\gamma{30}}$ is superior, the improvements are substantial (e.g., in Battleship-3 and Goofspiel-4).
On average, HS-DCFR(30) outperforms HS-DCFR(15) by $1.6$ OoM, and HS-PCFR$^+$(30) outperforms HS-PCFR$^+$(15) by $7.8$ OoM. Therefore, if only one schedule can be used, we recommend $\text{HS}_{\gamma{30}}$; otherwise, we recommend running both. 

\subsubsection{Ablation Studies}
\label{sec:ablt}

\begin{figure}[t]
\centering
    \subfloat{\includegraphics[width=0.429\columnwidth,trim={6 10 8 6},clip]{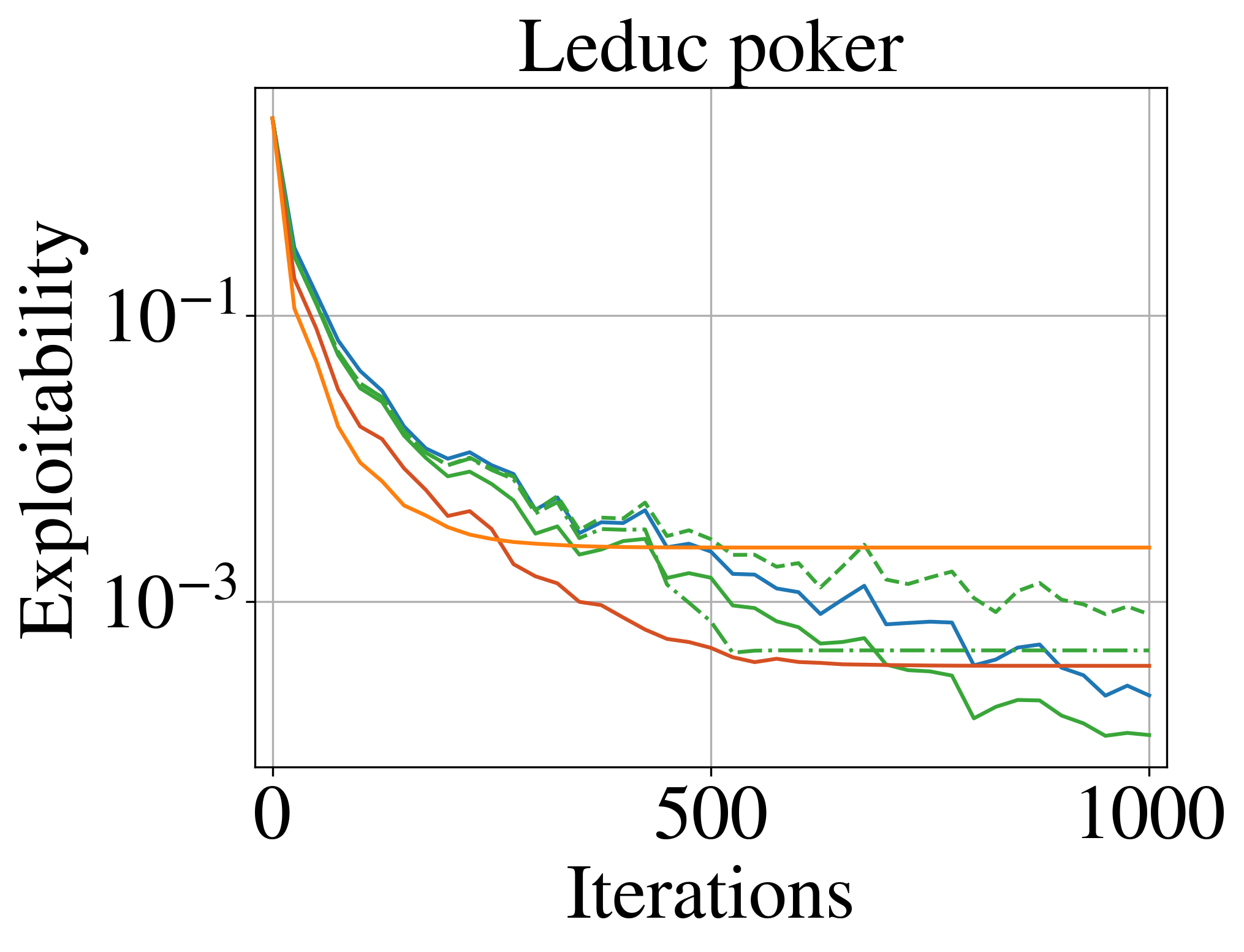}}
    \subfloat{\includegraphics[width=0.4\columnwidth,trim={6 10 8 6},clip]{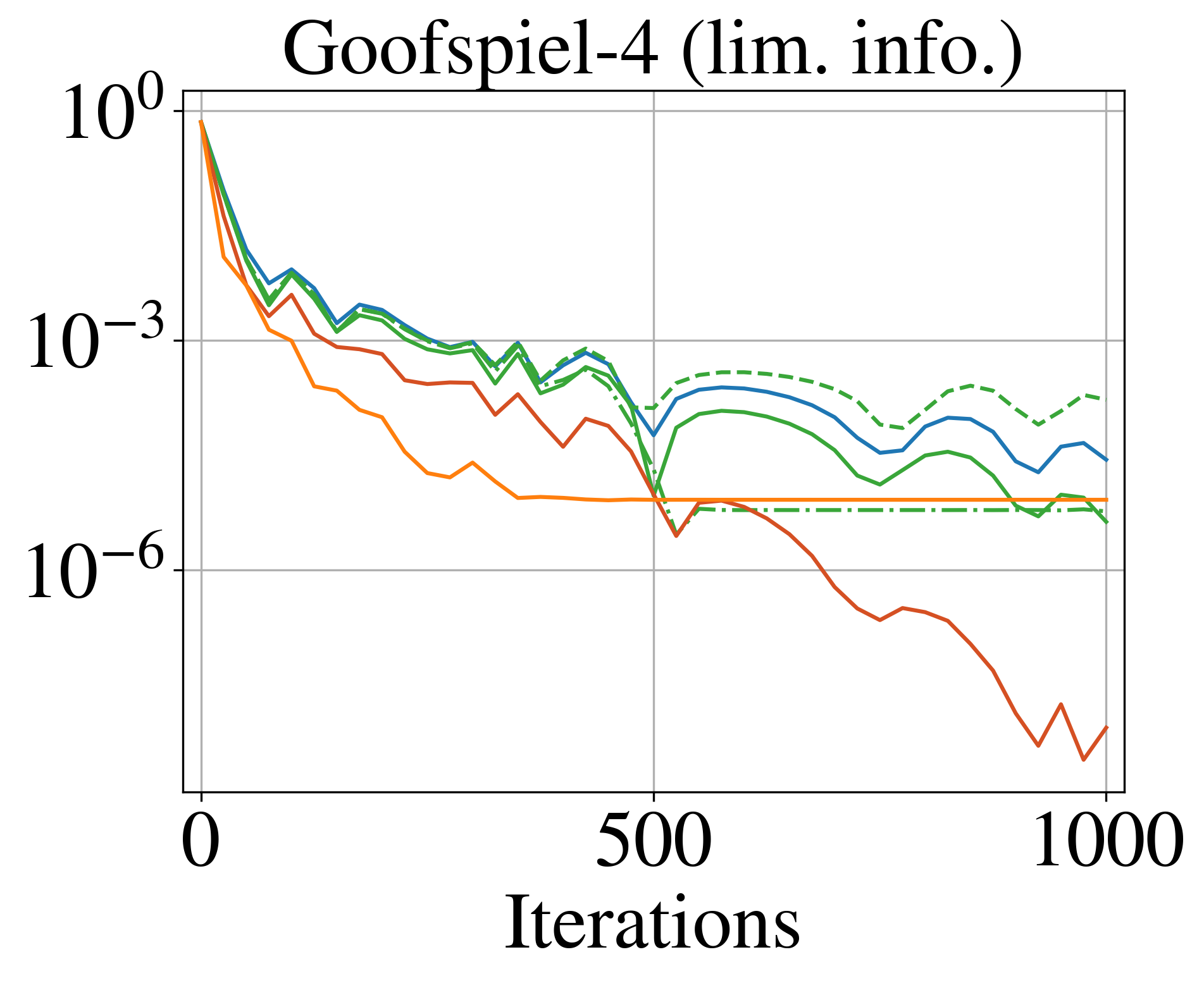}}
    \\
    \subfloat{\includegraphics[width=\linewidth,trim={15 125 15 125},clip]{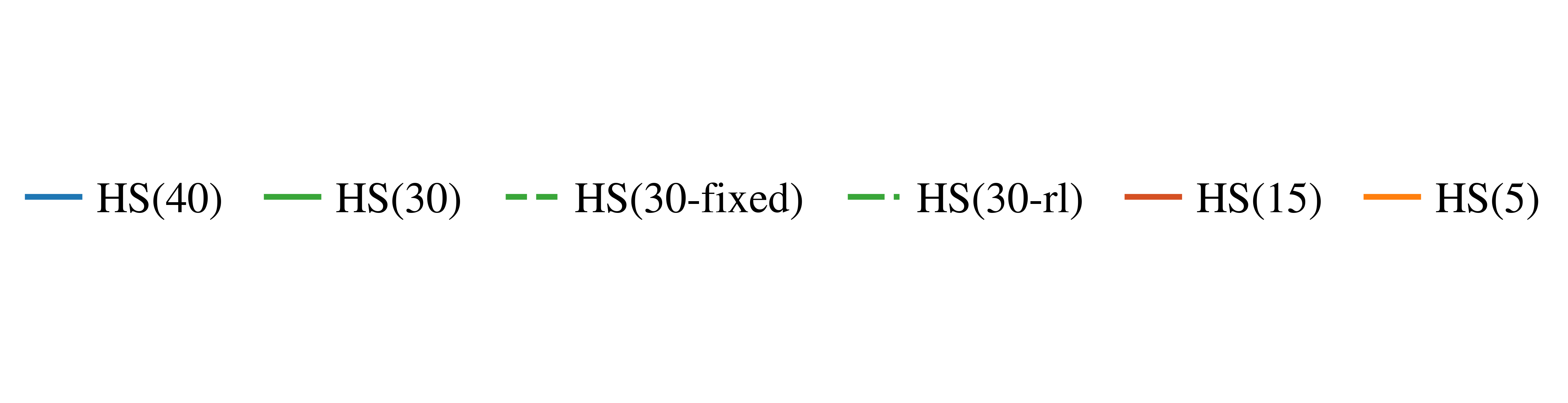}}
    \caption{Ablation studies on $\gamma$ using various HSs.}
    \label{fig:ablt}
    \vspace{-1mm}
\end{figure}

To better understand the strong empirical performance of HS-powered algorithms, we conduct ablation studies to analyze how different HSs of $\alpha$, $\beta$, and $\gamma$ in HS-powered algorithms influence performance. 
We consider HS-DCFR that is parameterized by ($\alpha$, $\beta$, $\gamma$) and fix $\alpha$ and $\beta$ according to the HSs proposed in Equation~\ref{eq:hs}. We then examine how the initial value of $\gamma$ (large or small) or its changing pattern (fixed, linear, or non-linear) across iterations contributes to the effectiveness of $\textsc{HS}_{\gamma}$. Figure~\ref{fig:ablt} shows the ablation results for $\gamma$, and similar studies for $\alpha$ and $\beta$ are provided in Appendix~\ref{apdx:ablt}.
Specifically, HS(40) and HS(5) follow the same changing pattern as HS(30) and HS(15) but use initial $\gamma$ values of $40$ and $5$, respectively. HS(30-fixed) uses a fixed $\gamma = 30$, while HS(30-rl) follows the $\gamma$ schedule of DDCFR. We evaluate these variants on both a poker game, Leduc poker, and a non-poker game, Goofspiel-4 (lim. info.).
Among the HSs of $\gamma$, HS(30) achieves the best performance in Leduc poker while HS(5) performs the worst. The non-linear changing scheme for $\gamma$ suggested by DDCFR plateaus early and the fixed scheme performs even worse. In Goofspiel-4 (lim. info.), HS(15) yields the best performance, whereas HS(30-fixed) performs the worst, and the non-linear scheme again plateaus early around iteration 500. Thus, combining a well-chosen initial $\gamma$ with a linear scheme results in strong performance.

\section{Conclusions}
\label{sec:concl}

In this work, we introduced \textit{Hyperparameter Schedules (HSs)}, a simple, training-free framework that dynamically adjusts discounting in CFR variants over time. Motivated by the observation that early strategies are less reliable than those from later iterations, HSs aggressively suppress the influence of early iterations and gradually increase trust in later updates.
With fewer than 15 lines of code changes to DCFR and PCFR$^+$, our resulting algorithms, HS-DCFR and HS-PCFR$^+$, establish new SoTA performance in both extensive-form and normal-form settings, often achieving orders of magnitude improvements in convergence rate.
We acknowledge that game-specific or game-class-specific tuning could further enhance the performance of HSs. 
Moreover, in games where the relative advantage of HSs diminishes as the game size increases, such tuning may help counteract this diminishing benefit. Nevertheless, the proposed HSs serve as a strong starting point for future research, as they already achieve SoTA performance across a wide range of games even without game-specific tuning. 

\section*{Acknowledgments}
This material is based on work supported by the Vannevar Bush Faculty Fellowship ONR N00014-23-1-2876, National Science Foundation grant RI-2312342, ARO award W911NF2210266, and NIH award A240108S001. Stephen McAleer was supported by a CRA Computing Innovation Fellow postdoctoral fellowship.

\bibliography{ref}

\clearpage
\appendix

\section{Proof of Theorem 3.1}
\label{apdx:proof}

The following proof\footnote{Note that, following a common simplification adopted in many prior works, we prove the exploitability bound under simultaneous updates. We refer readers to the work of Burch et al.~\cite{burch2019revisiting} for a proof showing how to recover the
exploitability bound under alternating updates.} is mainly based on rescaling inequalities over the ranges of \(\alpha\), \(\beta\), and \(\gamma\), and otherwise largely follows the proof of Theorem~1 in DDCFR~\cite{ddcfr}, which in turn follows the proof of Theorem~2 in DCFR~\cite{brown2019solving}. Here, we reiterate DDCFR’s proof of Theorem~1 with our modifications. In particular, this proof leverages the following lemmas from the existing literature.

\begin{lemma}~\cite[Lemma 5]{ddcfr} 
\label{lemma:1}
Assume that player $i$ conducts $T$ iterations of DDCFR. Then the weighted regret for the player is at most $\Delta\left|\mathcal{I}_i\right| \sqrt{|\mathcal{A}|} \sqrt{T}$ and the weighted average regret for the player is at most $\frac{8}{3} \Delta\left|\mathcal{I}_i\right| \sqrt{|\mathcal{A}|} / \sqrt{T}$.
\end{lemma}

\begin{lemma}~\cite[Lemma 1]{brown2019solving} 
\label{lemma:2}
Call a sequence $x_1, \ldots, x_T$ of bounded real values $BC$-plausible if $B>0, C \leq 0, \sum_{t=1}^{i} x_t \geq C$ for all $i$, and $\sum_{t=1}^T x_t \leq B$. For any $BC$-plausible sequence and any sequence of non-decreasing weights $w_t \geq 0, \sum_{t=1}^T\left(w_t x_t\right) \leq w_T(B-C)$. 
\end{lemma}

\textit{Proof}. By definition, the minimum instantaneous regret on any iteration is $-\Delta$. On iteration $t+1$, HS-DCFR scales negative regrets by $\frac{t^{\beta}}{t^{\beta}+1} \leq \frac{t^0}{t^0+1}=\frac{1}{2}$. Therefore, the regret for any action at any decision point is greater than $-2\Delta$. 

Let us consider the weighted sequence of iterates $\sigma^{\prime 1}, \ldots, \sigma^{\prime T}$, where $\sigma'^t$ is identical to $\sigma^t$ but assigned weight $w_{a, t}=\Pi_{i=t+1}^{T} \left(\frac{i-1}{i}\right)^{\gamma}$ rather than $w_t=\Pi_{i=t+1}^{T} \frac{(i-1)^{\alpha}}{(i-1)^{\alpha}+1}$. Let $R^{\prime t}(I, a)$ denote the regret of an action $a$ in an information set $I$ on iteration $t$ of the weighted sequence $\sigma^{\prime 1}, \ldots, \sigma^{\prime T}$. 

It follows from Lemma~\ref{lemma:1} that, for action $a$ in information set $I$ for player $i$, $R^T(I, a) \leq \frac{8}{3} \Delta \sqrt{|A|} \sqrt{T}$. Since $w_{a, t}$ is an increasing sequence, we can apply Lemma~\ref{lemma:2} using weight $w_{a, t}$ for iterations $t$ with $B=\frac{8}{3} \Delta \sqrt{|A|} \sqrt{T}$ and $C=-2 \Delta$, which gives us 
\begin{equation}
\label{eq:regret}
    R^{\prime T}(I, a) \leq \frac{8}{3} \Delta \sqrt{|A|} \sqrt{T}+2 \Delta.
\end{equation}
Given the upper bound on $\gamma$, the weighted sum satisfies
\begin{equation}
\label{eq:weighted_sum}
\resizebox{0.9\columnwidth}{!}{$
    \sum_{t=1}^T w_{a, t} \geq \sum_{t=1}^T \left(\prod_{i=t+1}^{T} \left(\frac{i-1}{i}\right)^U\right) = \sum_{t=1}^T \left(\frac{t}{T}\right)^U \geq \frac{T}{U+1}.
$}
\end{equation}
The weighted average regret therefore is
\begin{equation}
\resizebox{0.9\columnwidth}{!}{$
R_i^{\prime w, T} = \max_{a\in\mathcal{A}} \frac{R^{\prime T}(I, a)}{\sum_{t=1}^T w_{a, t}} \leq (U+1) \Delta\left(\frac{8}{3} \sqrt{|A|}+\frac{2}{\sqrt{T}}\right) / \sqrt{T}.
$}
\end{equation}
Applying Theorem $3$ in CFR~\cite{zinkevich2007regret}, the overall weighted average regret for player $i$ is at most $(U+1) \Delta|\mathcal{I}_i|\left(\frac{8}{3} \sqrt{|A|}+\frac{2}{\sqrt{T}}\right) / \sqrt{T}$. As $|\mathcal{I}_1| + |\mathcal{I}_2| = |\mathcal{I}|$, we can conclude that the weighted average strategies form a $(U+1) \Delta|\mathcal{I}|\left(\frac{8}{3} \sqrt{|A|}+\frac{2}{\sqrt{T}}\right) / \sqrt{T}$ -Nash equilibrium after $T$ iterations. \qed

\section{Proof of Theorem 3.2}
\label{apdx:proof_thm2}

\textit{Proof}. By Theorem~3 in PCFR$^+$~\cite{farina2021faster}, PRM$^+$ guarantees a regret bound for all $T$ that in the worst case implies
\begin{equation}
R^{\prime T}(I,a) \le O(\sqrt{T}),
\end{equation}
where $R^{\prime T}(I,a)$ is defined as in the proof of Theorem~\ref{thm:hsdcfr}.
This proof follows the proof of Theorem~\ref{thm:hsdcfr} by replacing Equation~\ref{eq:regret} with the bound above. 
By Equation~\ref{eq:weighted_sum}, the weighted average regret therefore is 
\begin{equation}
R_i^{\prime w, T} = \max_{a\in\mathcal{A}} \frac{R^{\prime T}(I, a)}{\sum_{t=1}^T w_{a, t}} \leq (U+1) O(1) / \sqrt{T}.
\end{equation}
Applying Theorem $3$ in CFR~\cite{zinkevich2007regret}, the overall weighted average regret for player $i$ is at most $(U+1) |\mathcal{I}_i|O(1) / \sqrt{T}$. As $|\mathcal{I}_1| + |\mathcal{I}_2| = |\mathcal{I}|$, we can conclude that the weighted average strategies form a $(U+1) |\mathcal{I}|O(1) / \sqrt{T}$ -Nash equilibrium after $T$ iterations. \qed

\section{Game Descriptions}
\label{apdx:game_desc}

\begin{table}[!htb]
    \caption{Size of the extensive-form benchmark games.}
    \label{tab:game_sizes}
    \begin{center}
    \begin{small}
    \begin{sc}
    \resizebox{\columnwidth}{!}{%
    \begin{tabular}{lccc}
        \toprule
        Game                        & \#Histories                   & \#Info. Sets          & \#Leaves \\
        \midrule
        Kuhn poker                  & 58                            & 12                    & 30                        \\
        Battleship-$2$              & $1.0\times 10^{4}$            & $3.3\times 10^{3}$    & $5.6\times 10^{3}$        \\
        Battleship-$3$              & $7.3\times 10^{5}$            & $8.1\times 10^{4}$    & $5.5\times 10^{5}$        \\
        Goofspiel-$3$ (li)          & 67                            & 16                    & 36                        \\
        Goofspiel-$4$               & $1.1\times 10^{3}$            & $2.7\times 10^{2}$    & $5.8\times 10^{2}$        \\
        Goofspiel-$4$ (li)          & $1.1\times 10^{3}$            & $1.6\times 10^{2}$    & $5.8\times 10^{2}$        \\
        Goofspiel-$5$               & $2.7\times 10^{4}$            & $3.3\times 10^{3}$    & $1.4\times 10^{4}$        \\
        Goofspiel-$5$ (li)          & $2.7\times 10^{4}$            & $2.1\times 10^{3}$    & $1.4\times 10^{4}$        \\
        Leduc poker                 & $9.5\times 10^{3}$            & $9.4\times 10^{2}$    & $5.5\times 10^{3}$        \\
        Big Leduc poker             & $6.2\times 10^{6}$            & $1.0\times 10^{5}$    & $4.0\times 10^{6}$        \\
        Liar's dice-$4$             & $8.2\times 10^{3}$            & $1.0\times 10^{3}$    & $4.1\times 10^{3}$        \\
        Liar's dice-$6$             & $2.9\times 10^{5}$            & $2.5\times 10^{4}$    & $1.5\times 10^{5}$        \\
        HUNL endgame-$1$            & $3.0\times 10^{8}$            & $5.2\times 10^{4}$    & $1.9\times 10^{8}$        \\
        HUNL endgame-$2$            & $3.5\times 10^{8}$            & $6.1\times 10^{4}$    & $2.3\times 10^{8}$        \\
        HUNL endgame-$3$            & $4.0\times 10^{8}$            & $6.9\times 10^{4}$    & $2.6\times 10^{8}$        \\
        HUNL endgame-$4$            & $3.5\times 10^{10}$           & $6.7\times 10^{6}$    & $2.2\times 10^{10}$       \\
        \bottomrule
    \end{tabular}%
    }
    \end{sc}
    \end{small}
    \end{center}
\end{table}

In this section, we offer a detailed description of each extensive-form game and its variants. We measure the sizes of the benchmark games across various metrics and report the results in Table~\ref{tab:game_sizes}. Within the table, \#Leaves denotes the number of terminal histories in the game tree. 

\textbf{Kuhn poker}~\cite{kuhn1950simplified} is a minimal poker game played with a three-card deck consisting of J, Q, and K. At the start of each hand, both players contribute one chip to the pot and are dealt a single private card. A single betting round then follows, during which players may choose to check, bet, call, or fold, subject to the constraint that each player may add at most one additional chip. If a fold occurs, the remaining player immediately wins the pot. Otherwise, after the betting round, the hand proceeds to a showdown in which both private cards are revealed and the player holding the higher-ranked card collects the entire pot.

\textbf{Battleship-$x$}~\cite{farina2019correlation} is derived from the classic Battleship board game. The game is played on a $2 \times x$ grid, where each player secretly selects a placement for a single $1 \times 2$ ship with an assigned value of $2$. Players then alternate turns attempting to locate the opponent's ship by targeting grid locations, with at most three shots per player. A ship is destroyed once all of its occupied cells have been hit, at which point the game ends immediately. Payoffs are defined as the difference between the value of the opponent's sunk ship and the value of the player's own sunk ship, if any.

\textbf{Goofspiel-$x$}~\cite{ross1971goofspiel} is a bidding game in which players allocate limited resources (i.e., cards) across a sequence of rounds. Each player holds a private hand of $x$ cards labeled from $1$ to $x$, each of which may be used exactly once. A separate deck of $x$ point cards, ordered from highest to lowest in our setting, specifies the point value revealed at the start of every round. Players then simultaneously submit a hidden bid by selecting a remaining card from their hand. The higher bid wins the point card, while ties result in the card being discarded. After all $x$ rounds, the player with the higher total point value is declared the winner. 
\textbf{Goofspiel (lim. info.)-$x$}~\cite{lanctot2009monte} is a limited-information variant of Goofspiel in which players do not observe their opponent's bids. Instead, both players submit their chosen cards to an umpire, who determines the outcome of the round and either awards the point card to the higher bid or discards it in the event of a tie.

\textbf{Leduc poker}~\cite{southey2005bayes} is a small poker game that extends Kuhn poker by introducing public information and multiple betting stages. The game is played with a six-card deck consisting of two suits and three ranks (J, Q, and K). Each player is dealt a single private card and participates in two betting rounds. The first betting round occurs before any public information is revealed and allows bets of up to two chips. A single public card is then dealt, followed by a second betting round with larger bet sizes of up to four chips. At showdown, a player whose private card matches the public card (i.e., forms a pair) wins the pot; if no pair is formed, the pot is awarded to the player with the higher-ranked private card.
We also include \textbf{Big Leduc poker} as a larger benchmark game, which extends Leduc poker by using a larger deck of twenty-four cards across twelve ranks and permitting up to six raises per betting round.

\textbf{Liar's dice-$x$}~\cite{lisy2015online} is a bidding game played with one $x$-sided die per player. At the start of the game, both players roll their die privately and observe the result. The game consists of a sequence of bids that specify a lower bound on the combined dice outcomes, expressed as a claim that at least $p$ dice show the number $q$. Players alternate turns by either increasing the claim (i.e., increasing $p$, $q$, or both) or challenging the previous bid by calling ``liar.'' A challenge immediately ends the game and triggers a reveal of both dice. If the challenged claim is false, the challenger wins; otherwise, the bidder wins.

\textbf{HUNL endgame-$x$} represents an endgame scenario in heads-up no-limit Texas hold 'em (HUNL) with a fixed public board and a specified pot size. The game state is characterized by the public community cards, the pot size, a conditional distribution over each player's private hands, and a betting abstraction. At the start of the endgame, private hands are sampled according to the conditional distributions, after which players take turns selecting actions that include folding, checking, calling, or betting using the available bet sizes. The game terminates either when a player folds or when both players proceed to a showdown. At showdown, the pot is awarded to the player holding the stronger hand, with ties resulting in a split pot. All HUNL endgames evaluated in this paper are generated by \textit{Libratus}~\cite{brown2018superhuman}, where HUNL endgame-4 corresponds to subgame2 from the collection of open-sourced Libratus subgames\footnote{\url{https://github.com/Sandholm-Lab/LibratusEndgames}}.

\renewcommand{\figscale}{0.25}
\begin{figure*}[!htb]
\centering
    \subfloat{\includegraphics[width=\figscale\textwidth]{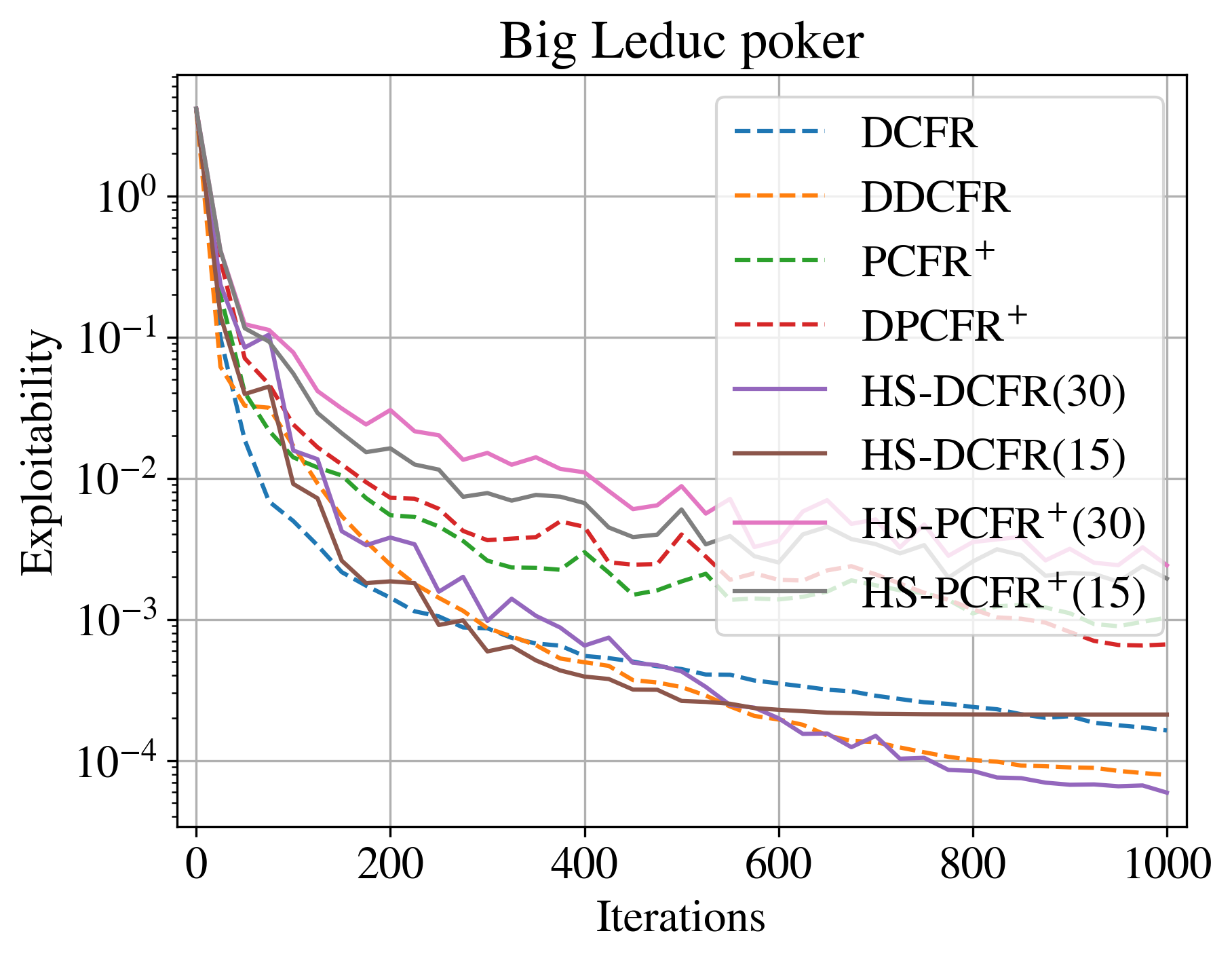}}
    \subfloat{\includegraphics[width=\figscale\textwidth]{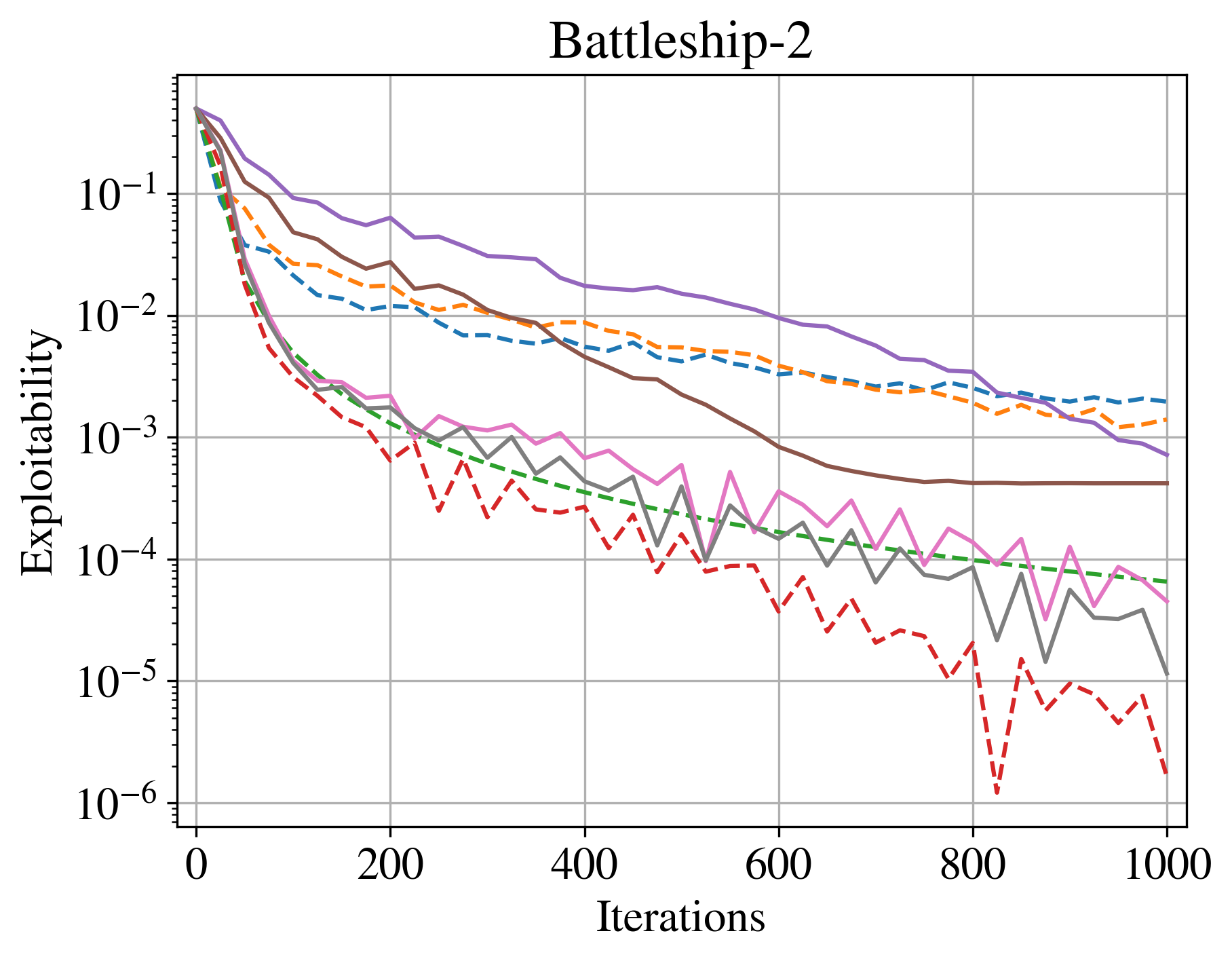}}
    \subfloat{\includegraphics[width=\figscale\textwidth]{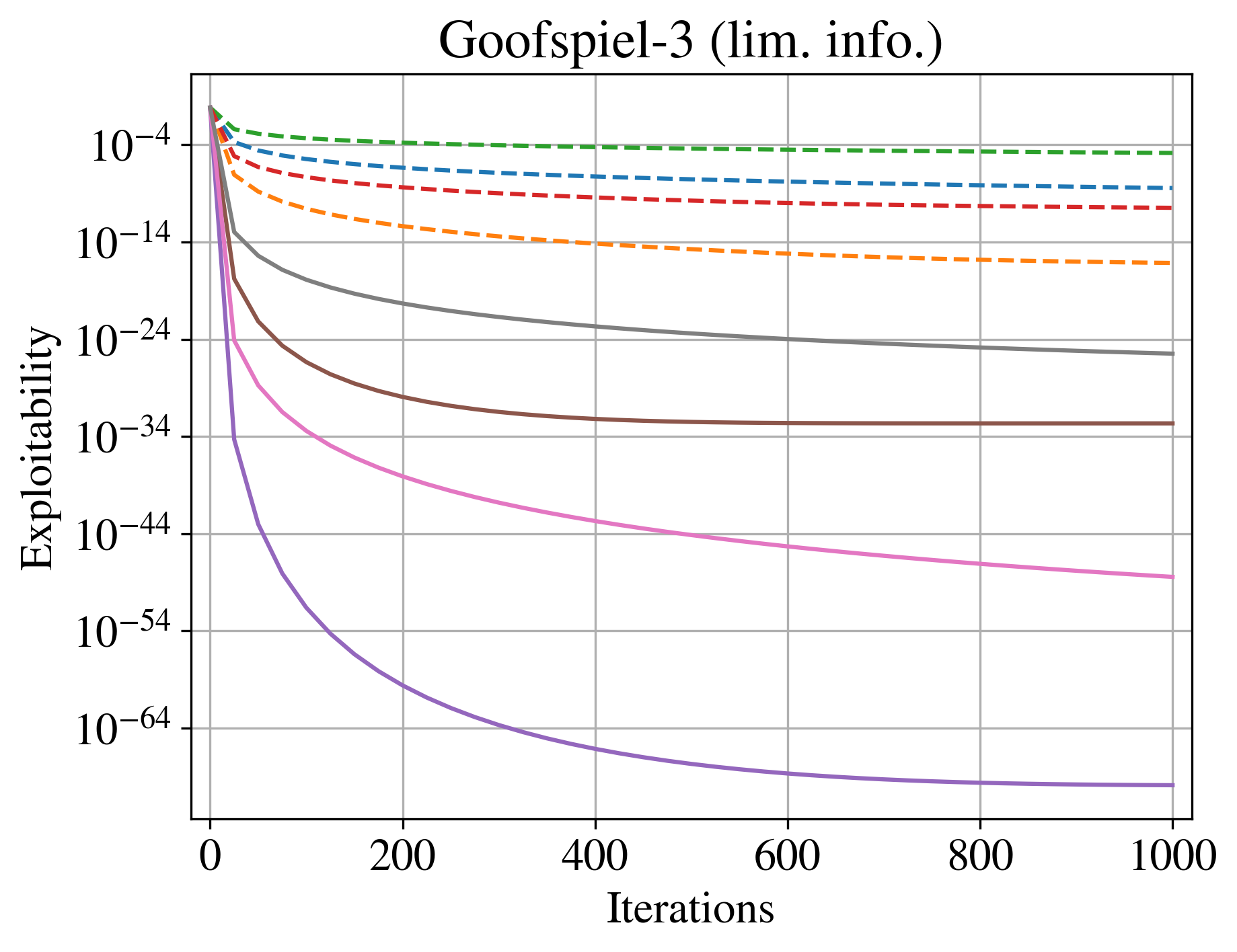}}
    \subfloat{\includegraphics[width=\figscale\textwidth]{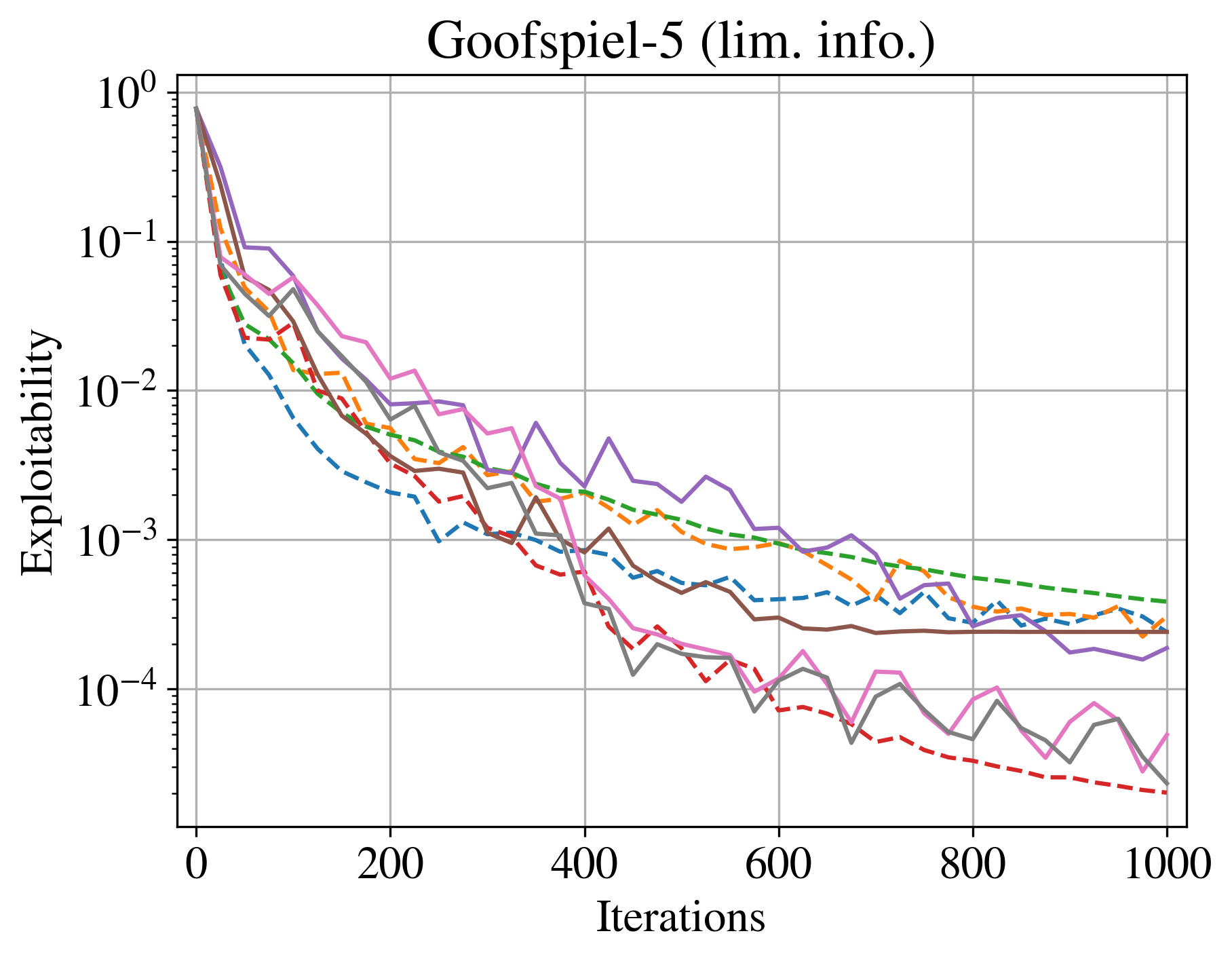}} \\
    \subfloat{\includegraphics[width=\figscale\textwidth]{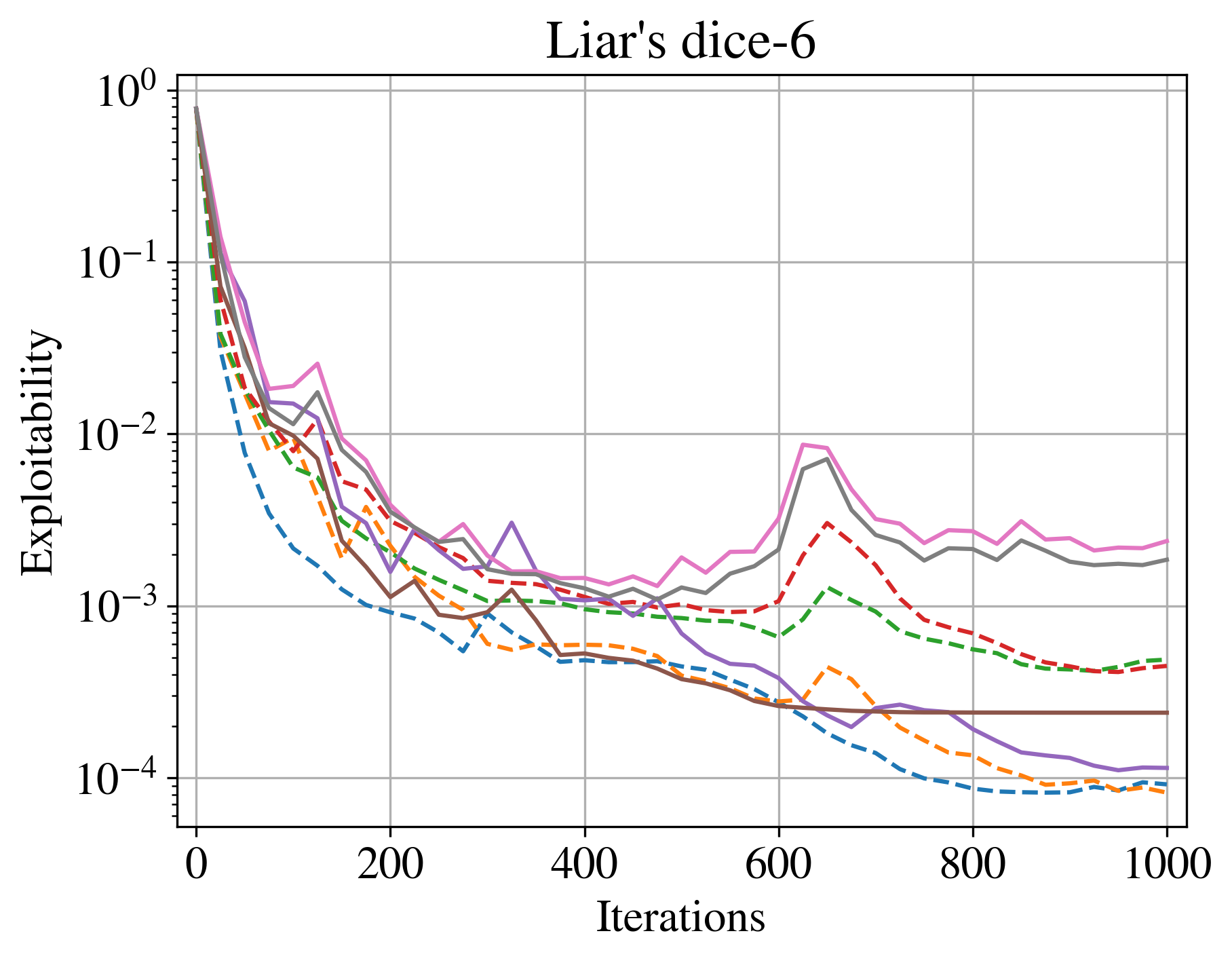}}
    \subfloat{\includegraphics[width=\figscale\textwidth]{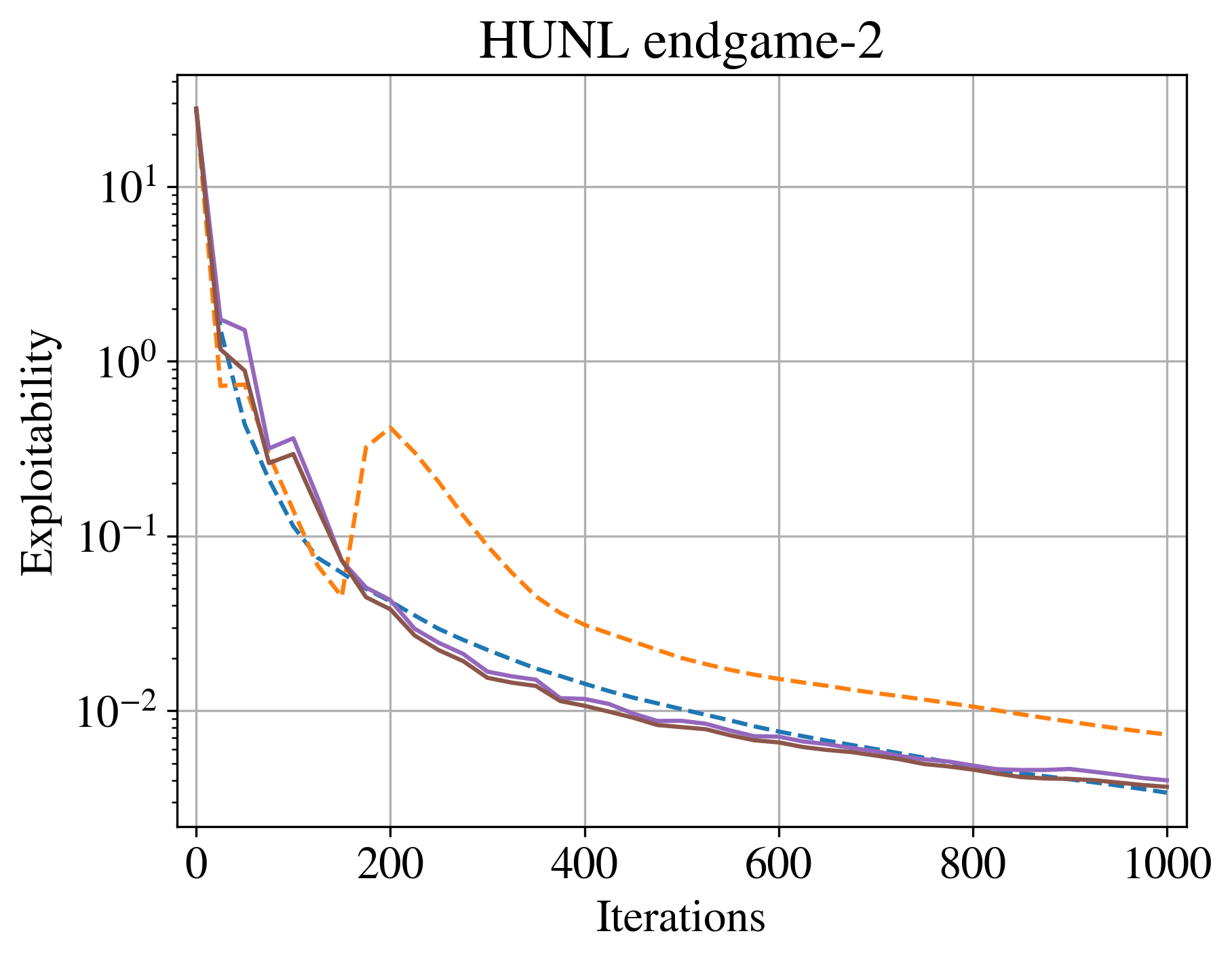}}
    \subfloat{\includegraphics[width=\figscale\textwidth]{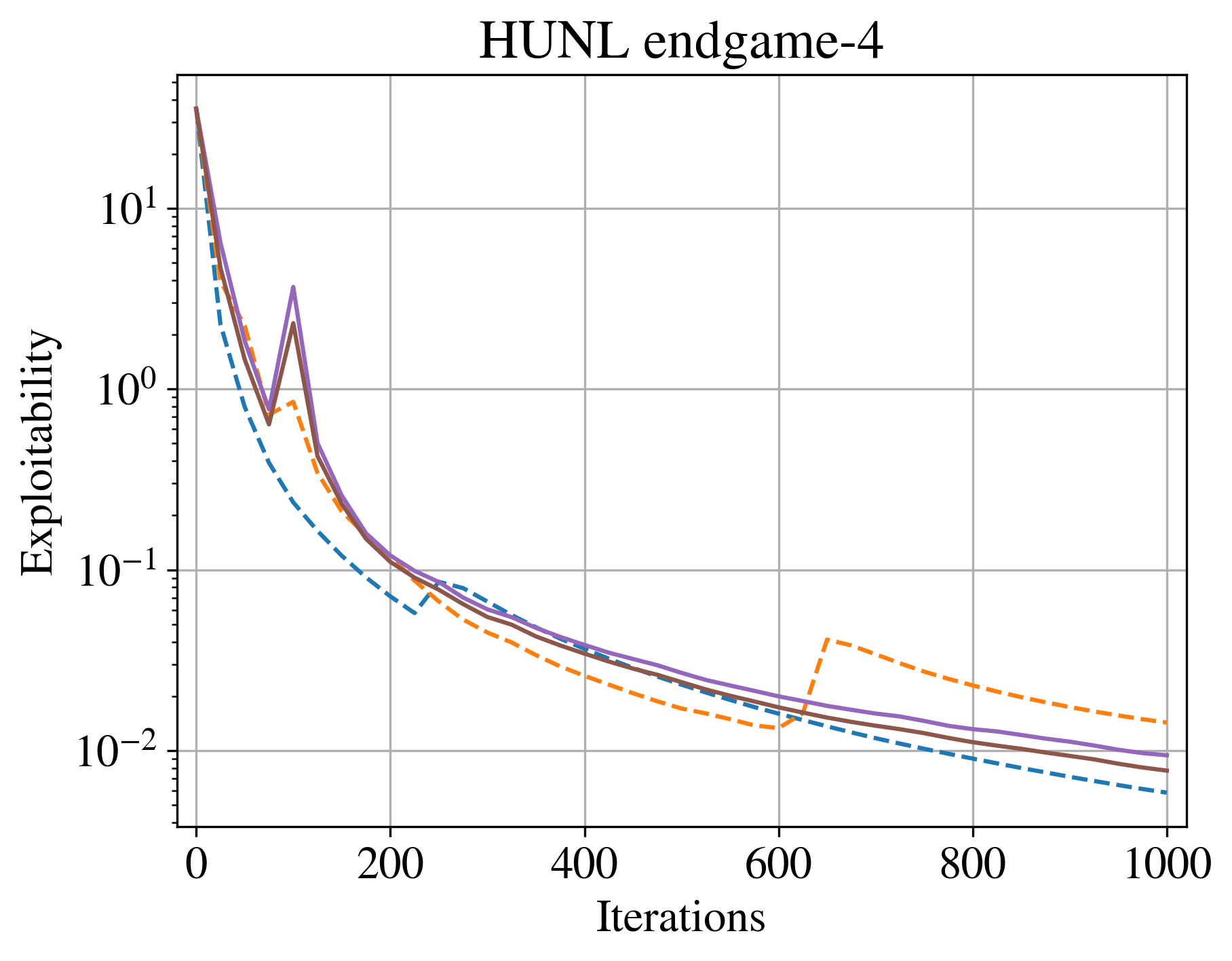}}
    \caption{Performance of HS-powered algorithms and prior SoTA algorithms on additional extensive-form games.}
    \label{fig:results_extra}
\end{figure*}

\section{Additional Experimental Results}
\label{apdx:results}

In Figure~\ref{fig:results_extra}, we present additional experimental results on the variants of the games discussed in the main body of the paper. In Battleship-2, HS-powered algorithms trail DPCFR$^+$ by $0.8$ OoM. However, in the extended Battleship game, they surpass DPCFR$^+$ by $3.2$ OoM, as shown in Figure~\ref{fig:results}. HS-powered algorithms also exhibit strong performance in Goofspiel-3 (lim. info.) and are on par with the prior SoTA in Big Leduc poker, Goofspiel-5 (lim. info.), and Liar's dice-6. In HUNL endgames, HS-DCFR consistently outperforms DDCFR.

\section{Additional Ablation Studies}
\label{apdx:ablt}

We conduct additional ablation studies to examine how different HSs for $\alpha$ and $\beta$ in HS-powered algorithms affect performance, with results reported in Figure~\ref{fig:ablt_extra}. 
Overall, performance is less sensitive to variations in $\alpha$ and $\beta$ than to changes in $\gamma$. In Leduc Poker, only HS(1-fixed) for $\alpha$ exhibits noticeably inferior performance relative to other variants, whereas all HS variants for $\beta$ achieve similar levels of exploitability by iteration 1,000. In Goofspiel-4 (lim. info.), the HSs for $\alpha$ exhibit greater performance divergence than in Leduc Poker. For HSs of $\beta$ in Goofspiel-4 (lim. info.), performance appears sensitive to the initial value, with an initial setting of $-1$ yielding improved results. An interesting observation is that employing a fixed scheme for $\alpha$ (i.e., HS(1-fixed)) results in the best performance in Goofspiel-4 (lim. info.) but the worst performance in Leduc Poker.

In Figure~\ref{fig:ablt_nc}, we show that simply ignoring early iterations is insufficient in both normal-form and extensive-form settings. For DCFR-NC, we follow the approach of Supremus~\cite{zarick2020unlocking} by discarding the contributions of the first third of iterations to the average strategy, setting their weights to zero. Across the four games evaluated, DCFR-NC eventually achieves exploitability close to that of DCFR by the final iteration; however, it consistently underperforms compared to the HS-DCFR variants. These results support our conclusion that disregarding early-iteration contributions is less effective than the more nuanced discounting schedules employed in our proposed approach.

\renewcommand{\figscale}{0.28}
\begin{figure*}[!htb]
\centering
    \subfloat{\includegraphics[width=0.27\textwidth]{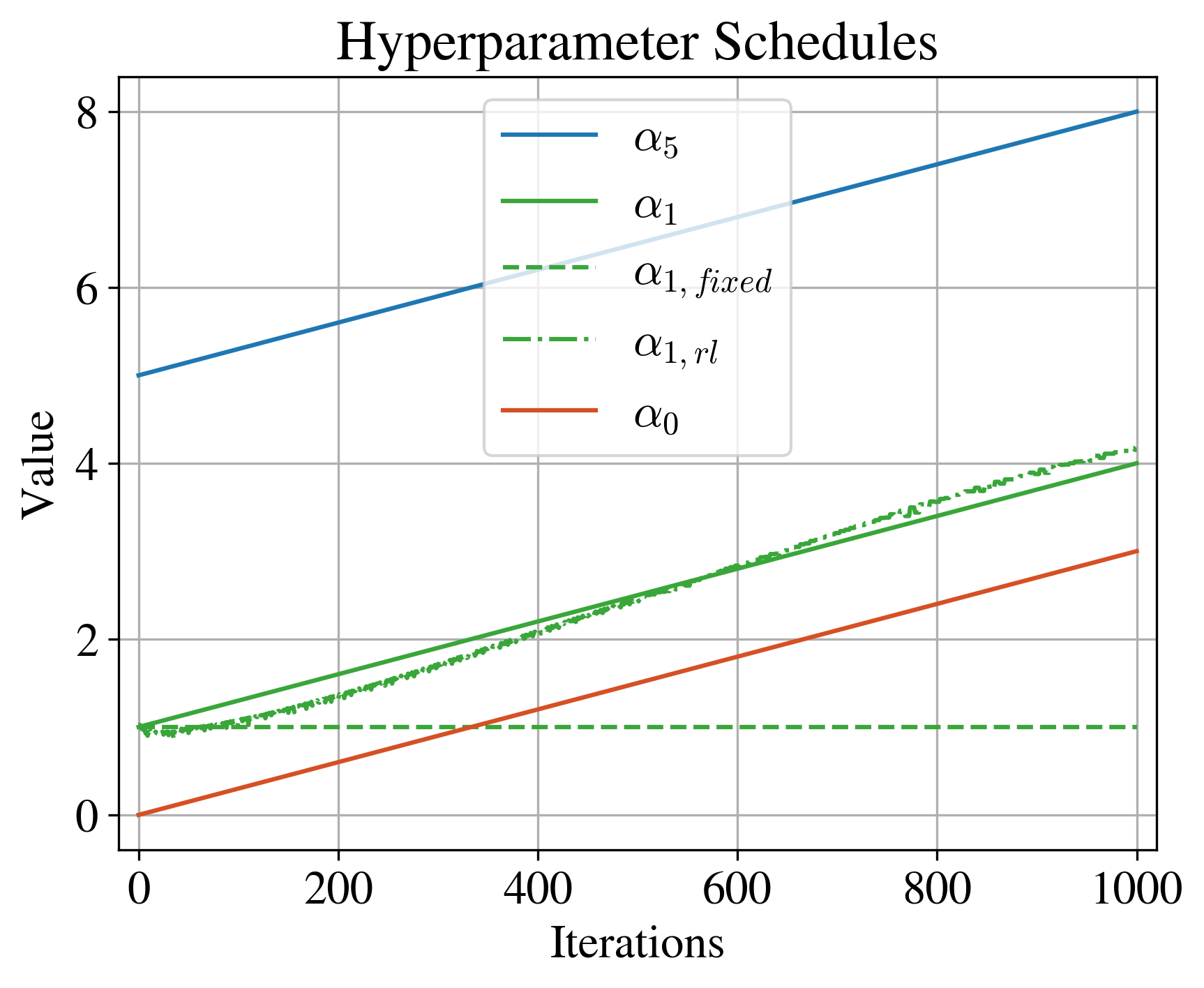}}
    \subfloat{\includegraphics[width=\figscale\textwidth]{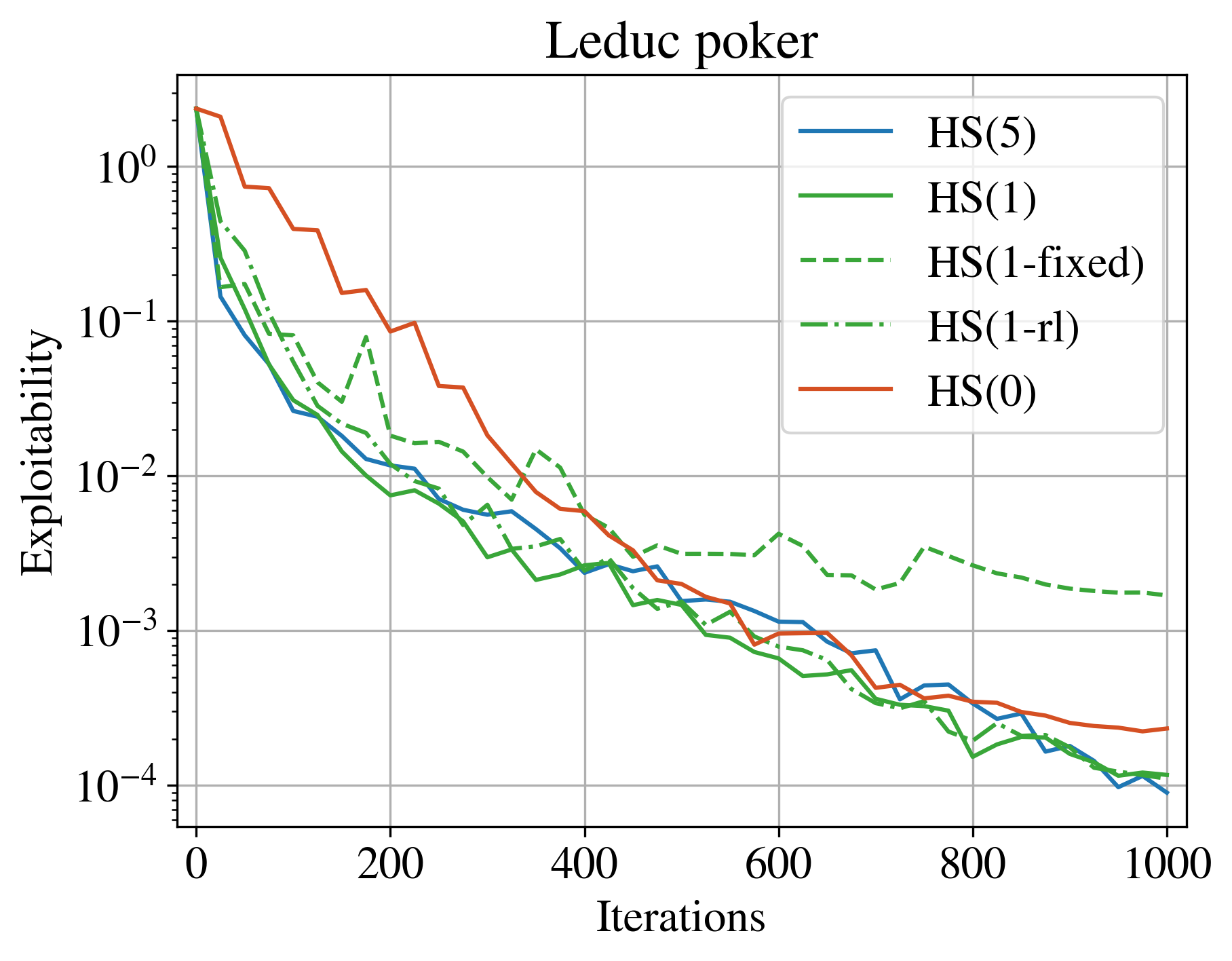}}
    \subfloat{\includegraphics[width=\figscale\textwidth]{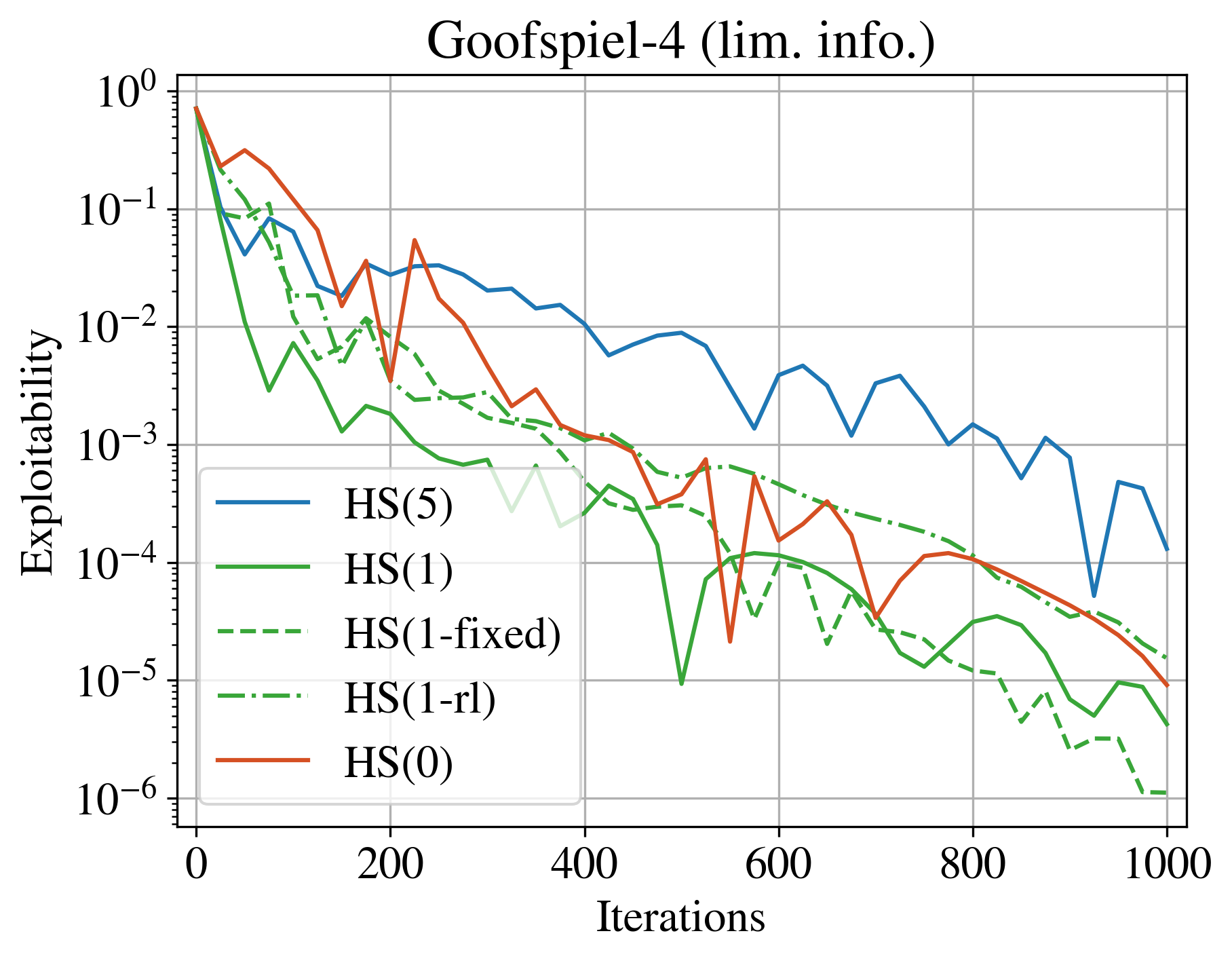}}
    \\ 
    \subfloat{\includegraphics[width=0.27\textwidth]{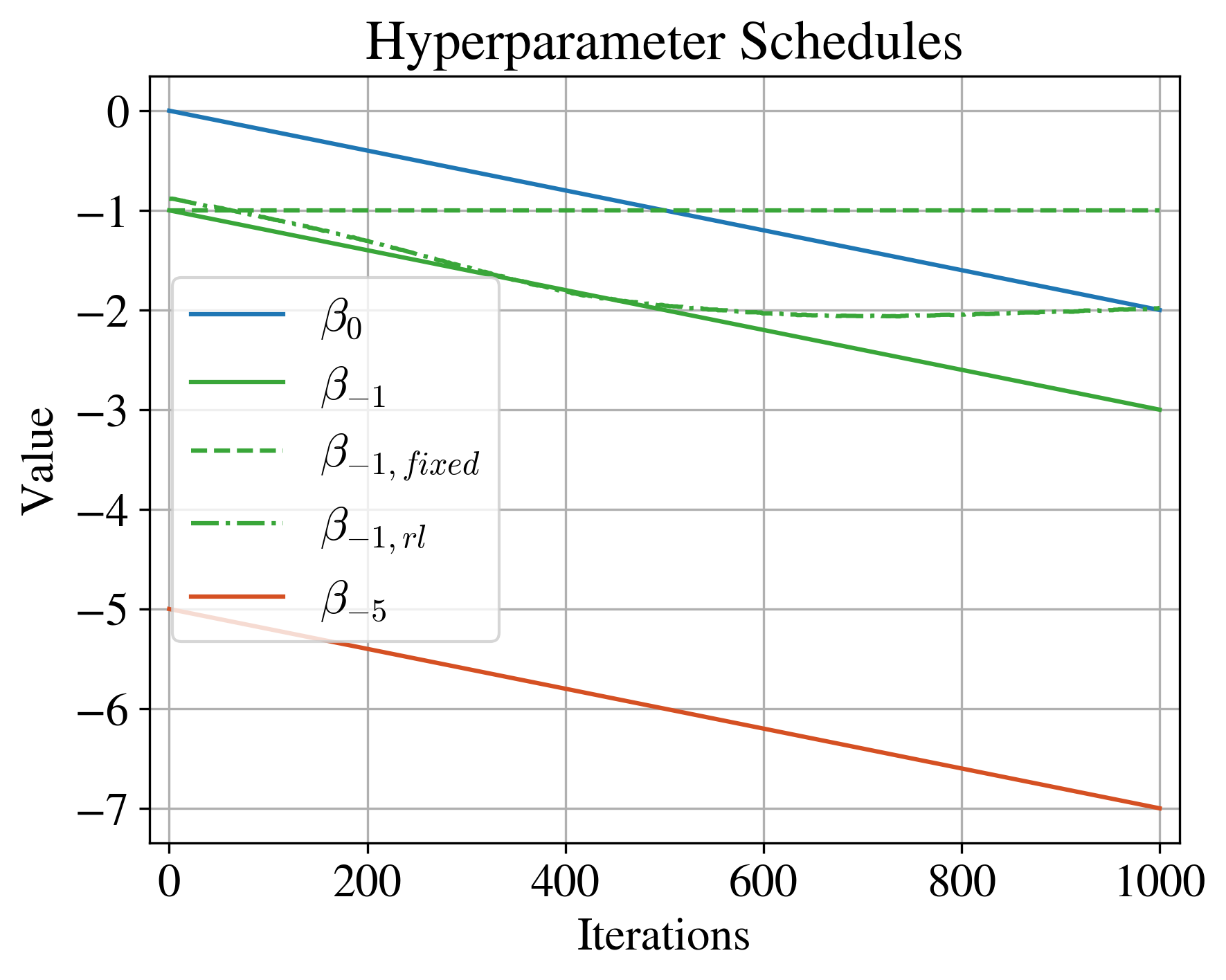}}
    \subfloat{\includegraphics[width=\figscale\textwidth]{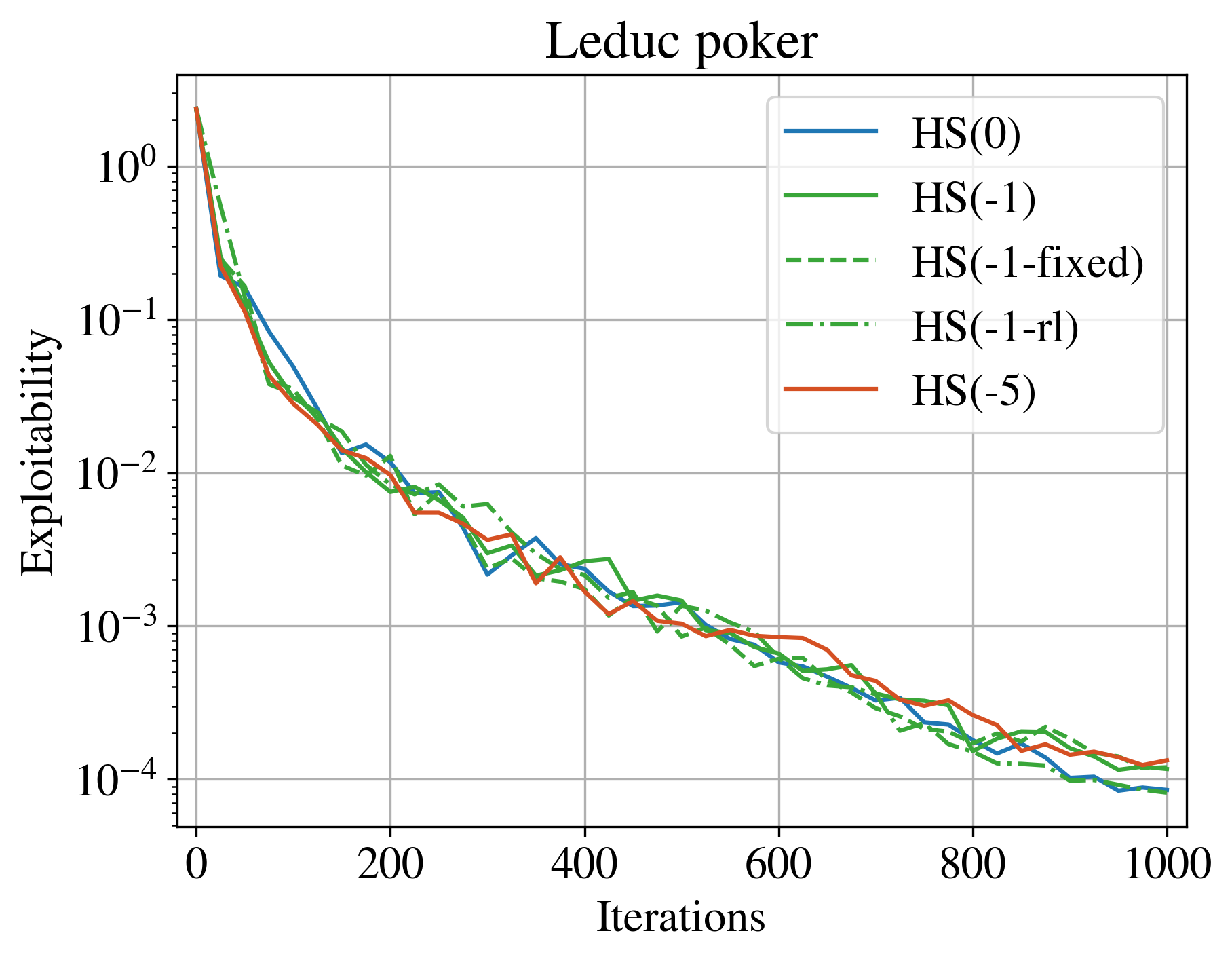}}
    \subfloat{\includegraphics[width=\figscale\textwidth]{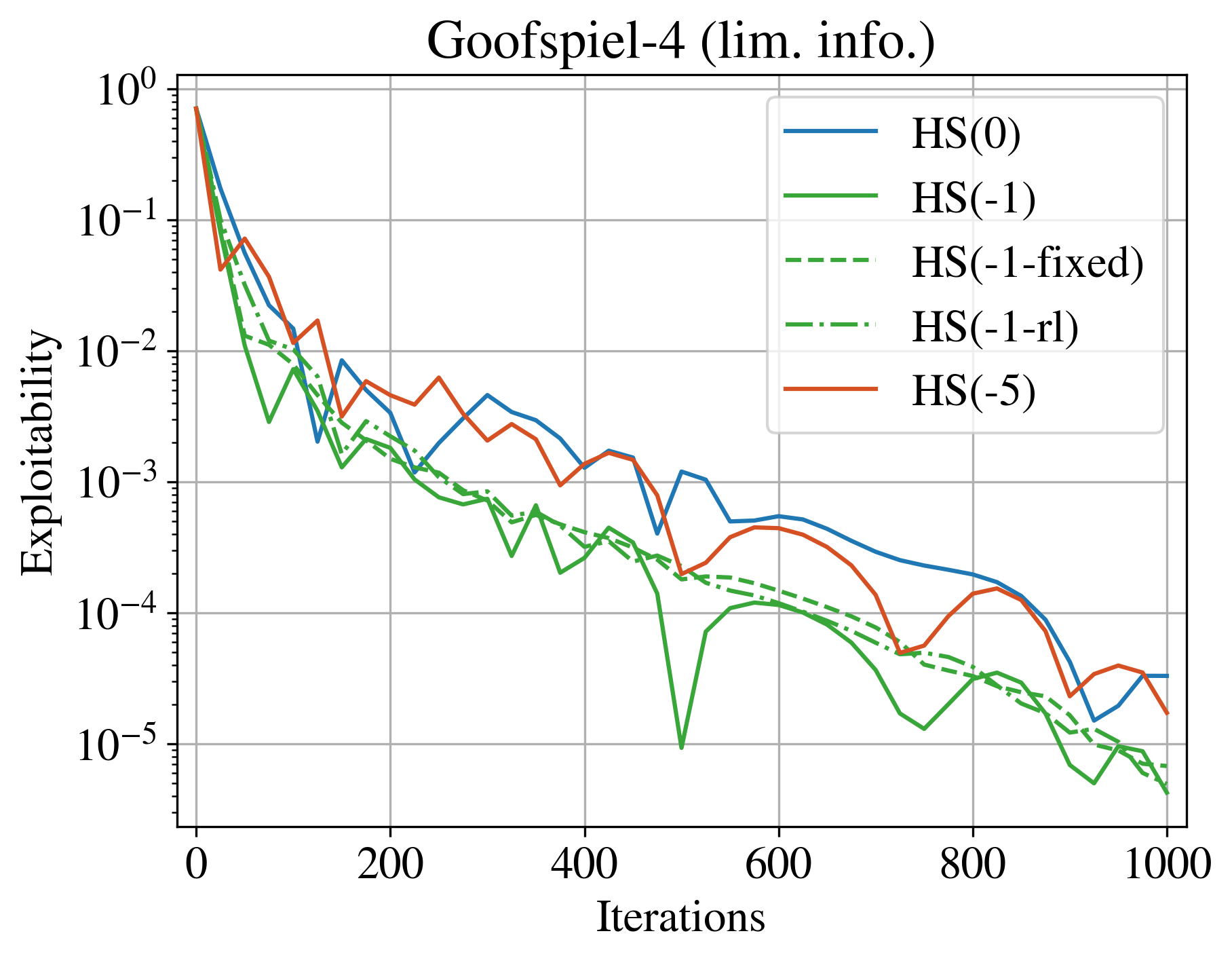}}
    \\ 
    \caption{Ablation studies on $\alpha$ and $\beta$ using various HSs.}
    \label{fig:ablt_extra}
\end{figure*}

\renewcommand{\figscale}{0.25}
\begin{figure*}[!htb]
\centering
    \subfloat{\includegraphics[width=\figscale\textwidth]{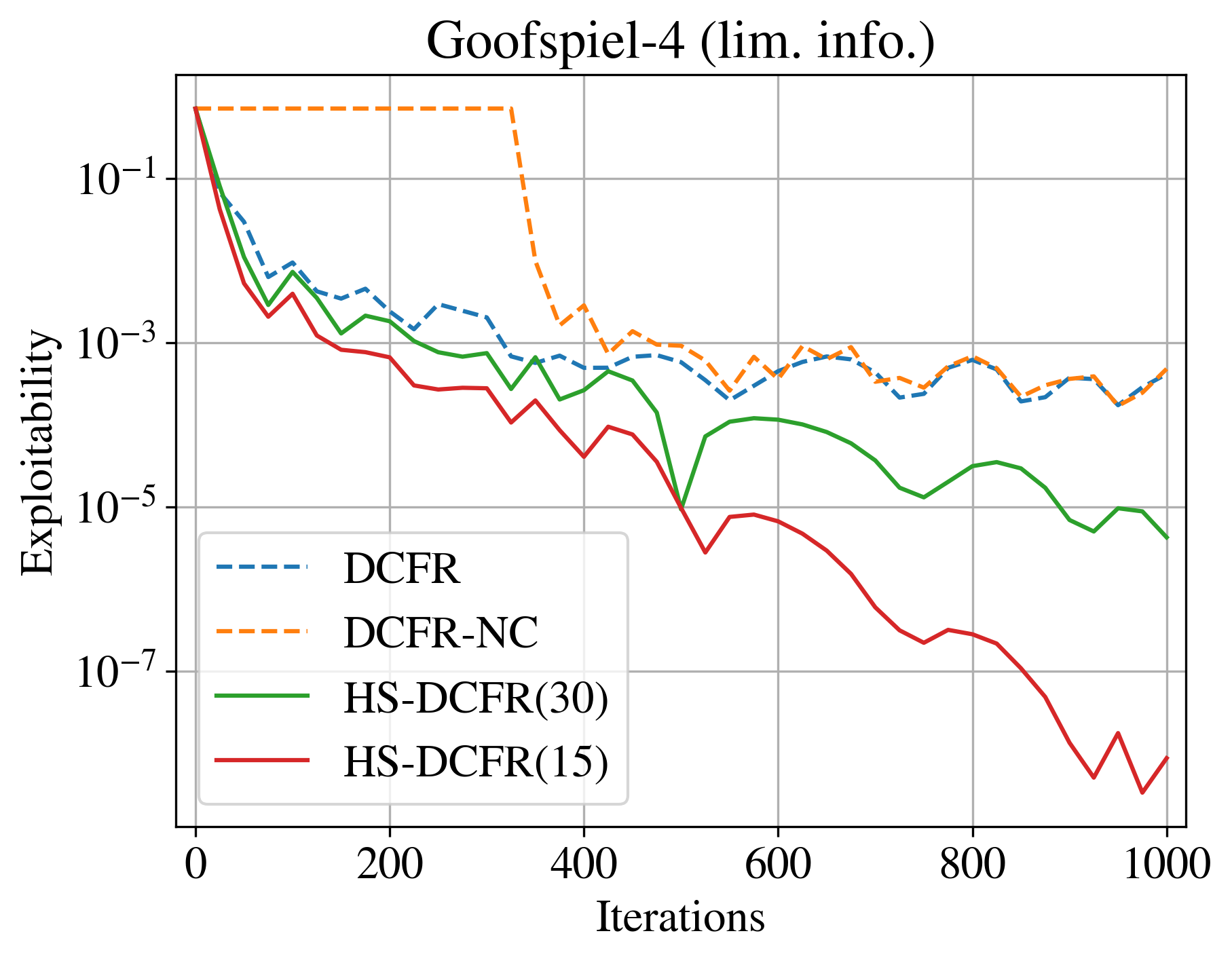}}
    \subfloat{\includegraphics[width=\figscale\textwidth]{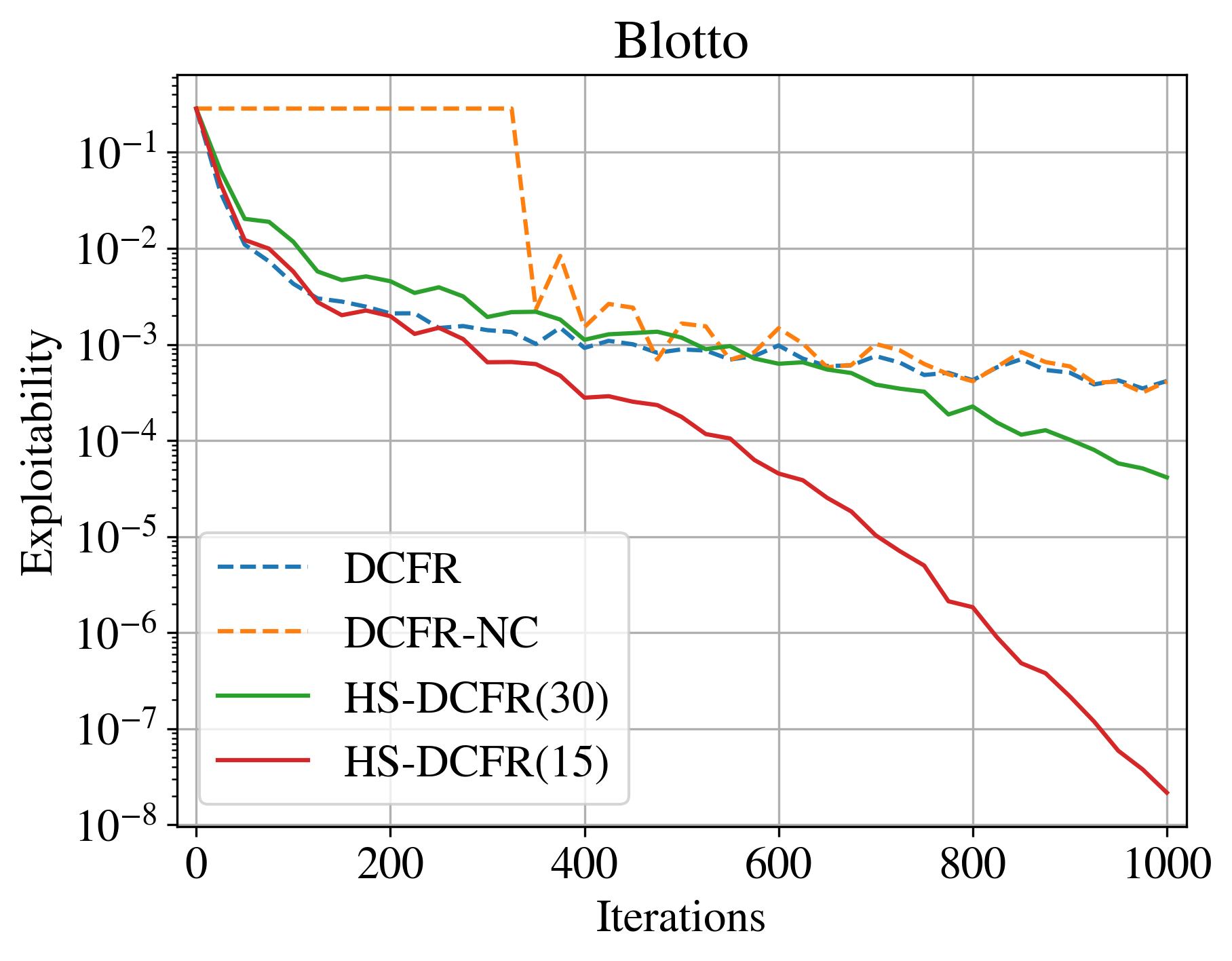}}
    \subfloat{\includegraphics[width=\figscale\textwidth]{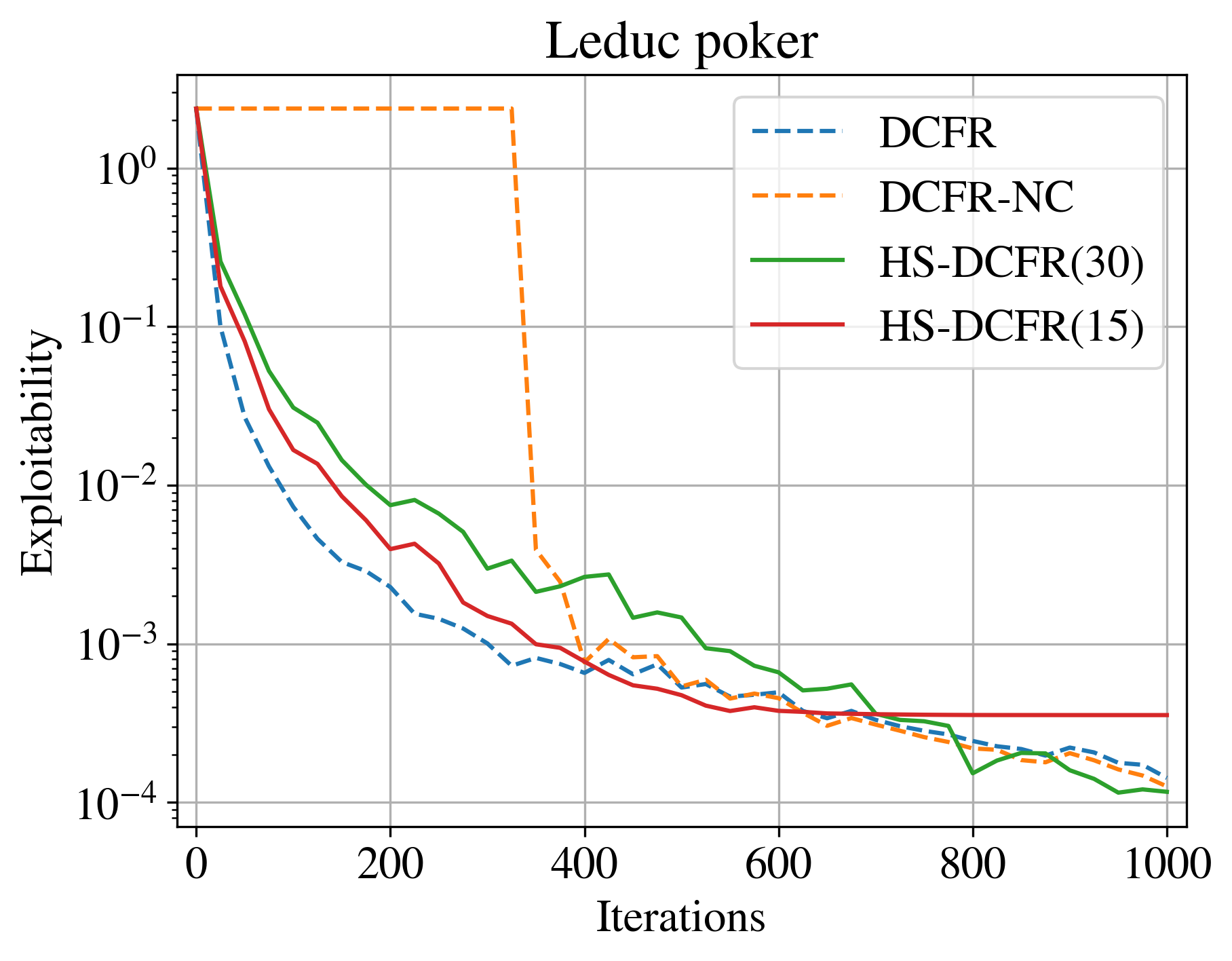}}
    \subfloat{\includegraphics[width=\figscale\textwidth]{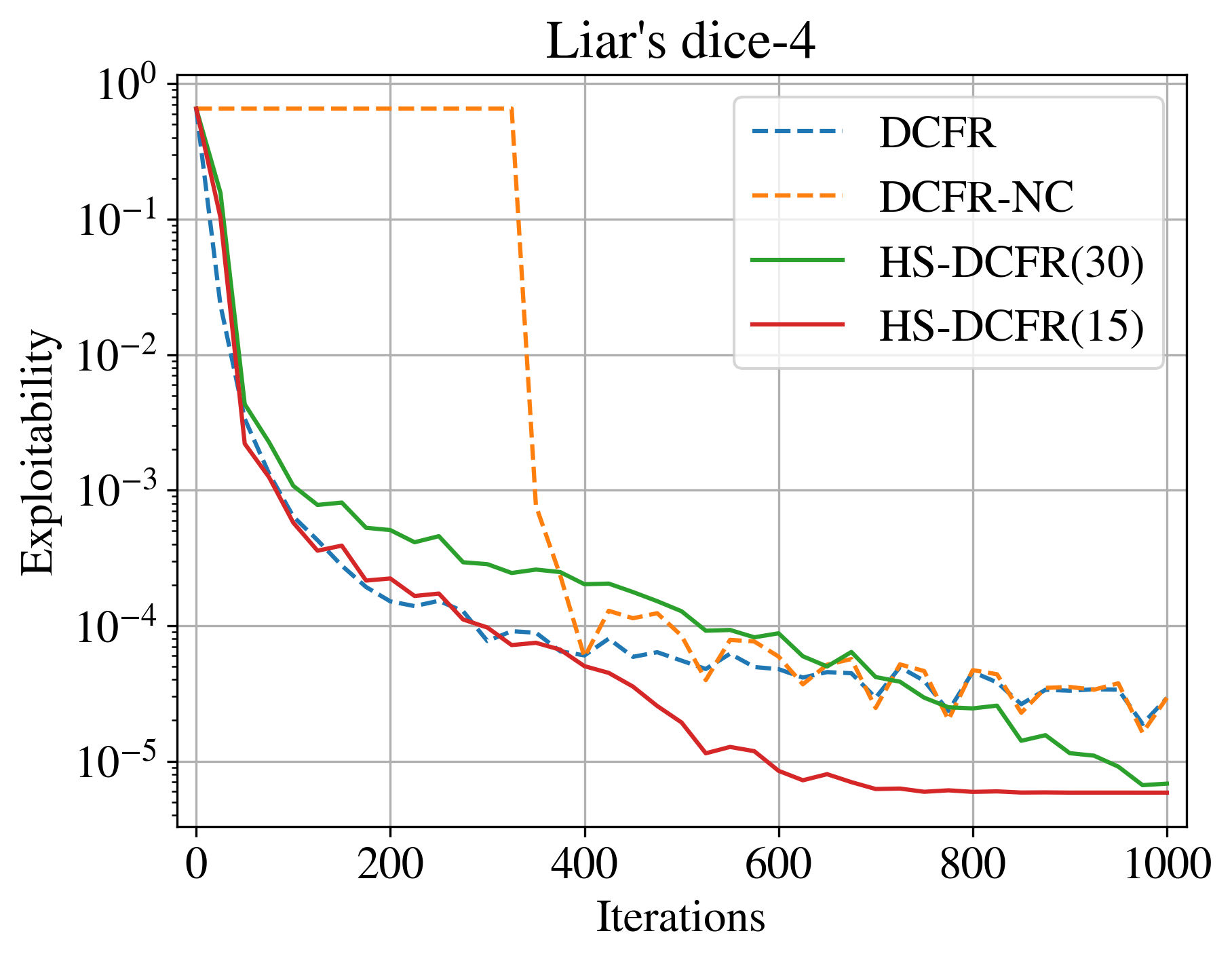}}
    \\ 
    \caption{Performance comparison of DCFR, DCFR-NC, and HS-DCFR.}
    \label{fig:ablt_nc}
\end{figure*}

\end{document}